\documentclass[11pt, reqno]{article}
\usepackage{jheppub}
\usepackage{graphicx}
\usepackage{amsmath}
\usepackage{amssymb}
\usepackage{amsthm}
\usepackage{tikz-network}
\usepackage{mathrsfs}
\usepackage{hyperref}
\usepackage{xcolor}
\usepackage{subfigure}
\usepackage{paratensor}
\usepackage{booktabs} 

\newcommand{\nn}{\nonumber \\}

\def\newline{{\hspace{15pt}}}

\def\abs#1{\left| #1\right|}

\def\bra#1{\left\langle #1\right|}
\def\ket#1{\left| #1\right\rangle}

\def\det{\mathop{\rm det}}

\def\eref#1{(\ref{#1})}

\def\ii{\text{i}}

\def\co{\,,}
\def\ed{\,.}

\title{\boldmath Spectral form factor of quadratic $R$-para-particle SYK model with Random Matrix Coupling  }

\author[a,b,c,d,e]{Tingfei Li}

\affiliation[a]{College of Physics Science and Technology, Hebei University, Baoding, 071002, China}
\affiliation[b]{Zhejiang Institute of Modern Physics, Zhejiang University, Hangzhou, 310027, P. R. China }
\affiliation[c]{College of Physics Science and Technology, Hebei University, Baoding 071002, China}
\affiliation[d]{Hebei Key Laboratory of High-precision Computation and Application of Quantum Field Theory, Baoding, 071002, China}
\affiliation[e]{Kavli Institute for Theoretical Sciences (KITS), University of Chinese Academy of Sciences, Beijing 100190, China}

\emailAdd{tfli@zju.edu.cn}
\abstract{This paper investigates the spectral form factor (SFF) of the quadratic $R$-para-particle Sachdev-Ye-Kitaev ($R$-PSYK$_2$) model with various random matrix ensemble couplings. We generalize previous work on Gaussian Unitary Ensemble (GUE) couplings to all three Gaussian ensembles (GUE, GOE, GSE) and three circular ensembles (CUE, COE, CSE). Through analytical and numerical methods, we establish precise correspondences between GUE and CUE results, demonstrating their SFFs satisfy $\mathcal{K}_{\text{GUE}}(2t) \approx \mathcal{K}_{\text{CUE}}(t)$ in the time regime $1 \ll t \ll N$. For the symplectic ensembles, we observe similar behavior with appropriate time rescaling, while we only provide the calculation method for the orthogonal ensembles. }

\begin{document}
	\maketitle
	\flushbottom

	\section{Introduction}
	\paragraph{Motivation} 
	It is well known that fundamental particles are classified into bosons and fermions. Their exchange statistics are governed by commutation and anticommutation relations of creation and annihilation operators. Beyond the three-dimensional fundamental particles, more statistical possibilities exist. For example, anyons in two-dimensional space \cite{1977NCimB_anyon,PhysRevLett.49.957,Wilczek1990,Nayak_2008,2008AnPhy} are important in topological quantum computation and quantum phases of matter. 
	Paraparticles \cite{PhysRev.90.270,Araki1961,Greenberg1965,Landshoff1967,Druehl1970,Robert1970,Baker2015} generalize ordinary exchange statistics and can be defined in any spatial dimension. First studied by Green in 1953 \cite{PhysRev.90.270}, they include bosons and fermions as special cases. Here, we mainly consider \textit{$R$-para-particles} defined by a \(R\)-matrix in their commutation relations \cite{Wang_2025,wang2025parastatisticssecretcommunicationchallenge,mekonnen2025invariancequantumpermutationsrules}. 
	Following standard notation, we represent the creation and annihilation operators for $R$-para-particles as
	\begin{align}
		\hat{\psi}_{i,a}^{+},\hat{\psi}_{i,a}^{-}
	\end{align}
	where $i = 1,2,\ldots,N$ denotes the site index and $a = 1,2,\ldots,m$ represents the flavor index. These operators obey generalized exchange statistics described by~\footnote{From now on, we write $\psi$ without the hat, since no confusion should arise.}
	\begin{equation}
		\begin{aligned}
			\hat{\psi}_{l,a}^-\hat{\psi}_{j,b}^+ &= \sum_{cd} R^{ac}_{bd} \hat{\psi}_{j,c}^+ \hat{\psi}_{l,d}^- + \delta_{ab} \delta_{lj}, \\
			\hat{\psi}_{l,a}^+ \hat{\psi}_{j,b}^+ &= \sum_{cd} R^{cd}_{ab} \hat{\psi}_{j,c}^+ \hat{\psi}_{l,d}^+, \\
			\hat{\psi}_{l,a}^- \hat{\psi}_{j,b}^- &= \sum_{cd} R^{ba}_{dc} \hat{\psi}_{j,c}^- \hat{\psi}_{l,d}^-\ed
		\end{aligned}
	\end{equation}
	In this formulation, each type of parastatistics is labeled by a four-index tensor $R^{ab}_{cd}$ satisfying  the constant
	Yang-Baxter equation (YBE) \cite{Turaev1988TheYE,10.1063/1.529485,etingof1998settheoreticalsolutionsquantumyangbaxter} %
	\begin{equation}\label{eq:YBE}
		\begin{tikzpicture}[baseline={([yshift=-.8ex]current bounding box.center)}, scale=0.5]
			\Rmatrix{0}{\AL}{R}
			\Rmatrix{0}{-\AL}{R}
			\node  at (-\AL,2.5*\AL) {\footnotesize $a$};
			\node  at (\AL,2.5*\AL) {\footnotesize $b$};
			\node  at (-\AL,-2.5*\AL) {\footnotesize $c$};
			\node  at (\AL,-2.5*\AL) {\footnotesize $d$};
		\end{tikzpicture}=
		\begin{tikzpicture}[baseline={([yshift=-.8ex]current bounding box.center)}, scale=0.5]
			\draw[thick] (-\AL,-2*\AL) -- (-\AL,2*\AL);
			\draw[thick] (\AL,-2*\AL) -- (\AL,2*\AL);
			\node  at (-\AL,2.5*\AL) {\footnotesize $a$};
			\node  at (\AL,2.5*\AL) {\footnotesize $b$};
			\node  at (-\AL,-2.5*\AL) {\footnotesize $c$};
			\node  at (\AL,-2.5*\AL) {\footnotesize $d$};
			\node  at (-1.5*\AL,0*\AL) {\footnotesize $\delta$};
			\node  at (1.5*\AL,0*\AL) {\footnotesize $\delta$};
		\end{tikzpicture}~,~~~
		\begin{tikzpicture}[baseline={([yshift=-.8ex]current bounding box.center)}, scale=0.5]
			\Rmatrix{-\AL}{2*\AL}{R}
			\Rmatrix{\AL}{0}{R}
			\Rmatrix{-\AL}{-2*\AL}{R}
			\draw[thick] (-2*\AL,-\AL) -- (-2*\AL,\AL);
			\draw[thick] (2*\AL,\AL) -- (2*\AL,3*\AL);
			\draw[thick] (2*\AL,-\AL) -- (2*\AL,-3*\AL);
			\node  at (-2*\AL,3.5*\AL) {\footnotesize $a$};
			\node  at (0*\AL,3.5*\AL) {\footnotesize $b$};
			\node  at (2*\AL,3.5*\AL) {\footnotesize $c$};
			\node  at (-2*\AL,-3.7*\AL) {\footnotesize $d$};
			\node  at (0*\AL,-3.7*\AL) {\footnotesize $e$};
			\node  at (2*\AL,-3.7*\AL) {\footnotesize $f$};
		\end{tikzpicture}
		=\begin{tikzpicture}[baseline={([yshift=-.8ex]current bounding box.center)}, scale=0.5]
			\Rmatrix{\AL}{2*\AL}{R}
			\Rmatrix{-\AL}{0}{R}
			\Rmatrix{\AL}{-2*\AL}{R}
			\draw[thick] (2*\AL,-\AL) -- (2*\AL,\AL);
			\draw[thick] (-2*\AL,\AL) -- (-2*\AL,3*\AL);
			\draw[thick] (-2*\AL,-\AL) -- (-2*\AL,-3*\AL);
			\node  at (-2*\AL,3.5*\AL) {\footnotesize $a$};
			\node  at (0*\AL,3.5*\AL) {\footnotesize $b$};
			\node  at (2*\AL,3.5*\AL) {\footnotesize $c$};
			\node  at (-2*\AL,-3.7*\AL) {\footnotesize $d$};
			\node  at (0*\AL,-3.7*\AL) {\footnotesize $e$};
			\node  at (2*\AL,-3.7*\AL) {\footnotesize $f$};
		\end{tikzpicture},
	\end{equation}
	where $R^{ab}_{cd}=\!\!\begin{tikzpicture}[baseline={([yshift=-.6ex]current bounding box.center)}, scale=0.45]
		\Rmatrix{0}{0}{R}
		\node  at (-1.5*\AL,\AL) {\footnotesize $a$};
		\node  at (1.5*\AL,\AL) {\footnotesize $b$};
		\node  at (-1.5*\AL,-\AL) {\footnotesize $c$};
		\node  at (1.5*\AL,-\AL) {\footnotesize $d$};
	\end{tikzpicture}$,
	and a line segment represents a Kronecker $\delta$ function. To maintain the physical interpretation of $\psi_{i,a}^{\pm}$ as creation and annihilation operators, we impose the additional condition
	\begin{align}\label{eq:unitary-condition}
		\sum_{a,b}R_{cd}^{ab}\left(R_{ef}^{ab}\right)^{*}=\delta_{ce}\delta_{df}\co	
	\end{align}
	ensuring that $\psi_{i,a}^+ = (\psi_{i,a}^-)^\dagger$. We emphasize that the special cases $R^{ab}_{cd} = \pm \delta_{ad}\delta_{bc}$ recover standard bosonic ($+$) and fermionic ($-$) statistics. 
	
	Recently, the SYK (Sachdev-Ye-Kitaev) model \cite{Kitaev-talks,Kitaev2015,Polchinski_2016,Maldacena_2016,Jevicki_2016,jevicki2016bilocalholographysykmodel,Kitaev:2017awl}---a system of fermions with random $q$-body interactions---has been widely studied for its connections to quantum chaos \cite{Maldacena_2016chaos,Parker:2018yvk,Gu:2016oyy,Roberts:2018mnp}, black holes \cite{Maldacena:2018lmt,Maldacena:2017axo,Nayak:2018qej}, holography \cite{Saad:2018bqo,Sarosi:2017ykf,Jevicki:2016bwu,Gross:2017hcz}, and quantum matter \cite{Davison:2016ngz,Hartnoll:2016apf,Song:2017pfw,Chowdhury:2021qpy}. We explore a variant using $R$-para-particles ($R$-PSYK), where nontrivial exchange statistics complicate the standard path integral approach. In \cite{Li:2025kpz}, we mainly study the $q=2$ SYK model constructed by $R$-para-particles, which we refer to as $R$-PSYK$_2$, whose Hamiltonian is quadratic with disordered coupling $h_{ij}$ being random Hermitian matrices  drawn from the Gaussian Unitary Ensemble (GUE). Like other models with only quadratic interactions, the $R$-PSYK$_2$ model can be diagonalized easily so that its spectrum is just the collection of non-interacting particles. However, such simplicity does not mean triviality. The coupling disorder can lead to chaotic behavior in the SFF of the model after taking ensemble average. The model exhibits an exponential ramp \cite{Lau_2019,Liao_2020,Winer_2020,Chris_Lau_2021,legramandi2024manybodyspectraltransitionslens} in its SFF dynamics: $\mathcal{K}(t) \sim e^{C_0 t}$ for $(N/C_0)^{2/5}\ll t \ll N$. The theoretical analysis implies that the growth rate $C_0$ tends toward a constant or infinity as $N\to \infty$, depending on the specific statistics of the model. However, a direct numerical comparison with the large $N$ theoretical predictions is lacking due to computational constraints and numerical errors. Recently, a SYK$_2$-like model with couplings related to the Circular Unitary Ensemble (CUE) was studied in \cite{Michael_2024}, where exact results for SFF were derived. Since CUE and GUE share the same cluster functions in the large $N$ limit, we expect the results in \cite{Michael_2024} to serve as a benchmark. In addition to CUE, we also want to check whether a similar phenomenon occurs for other ensembles, so we consider extending the model to four additional matrix ensembles: the Gaussian Orthogonal Ensemble (GOE), the Gaussian Symplectic Ensemble (GSE), the Circular Orthogonal Ensemble (COE), and the Circular Symplectic Ensemble (CSE).

	\begin{table}[ht]\label{table:six-cases}
		\centering
		\vspace{0.5cm}
		\begin{tabular}{l|ll} 
			\toprule
			& \textbf{$R$-para-fermions} & \textbf{Dual $R$-para-bosons} \\
			\midrule
			Ordinary & $(1+x)^m$ & $(1-x)^{-m}$ \\
			Example A & $1+mx$ & $(1-mx)^{-1}$ \\
			Example B & $1+mx+x^2$ & $(1-mx+x^2)^{-1}$ \\
			\bottomrule
		\end{tabular}
		\caption{The single-mode partition functions of the three cases of $R$-para-fermions and their dual $R$-para-bosons. The function $z_R(x) = 1 + m x$ arises when $R^{ab}_{cd} = -\delta_{ac} \delta_{bd}$, while $z_R(x) = 1 + m x + x^2$ corresponds to $R^{ab}_{cd} = \lambda_{ab} \xi_{cd} - \delta_{ac} \delta_{bd}$. Here, the matrices $\lambda$ and $\xi$ satisfy the conditions $\lambda \xi \lambda^T \xi^T = \mathbf{1}$ and $\operatorname{Tr}(\lambda \xi^T) = 2$. Note that the unitary condition in Eq.~\eref{eq:unitary-condition} cannot be satisfied for Example B with \( m \ge 3 \) for any \(\lambda, \xi\). As a result, we never have \(\psi_{i,a}^+ = (\psi_{i,a}^-)^\dagger\) in this case. However, since the partition function and SFF only depend  on \(d_n\), the physical results presented here remain valid. }
	\end{table}

	\paragraph{The model} 
	The $R$-PSYK$_2$ model in \cite{Li:2025kpz} is defined by a disordered Hamiltonian
	\begin{align}\label{eq:original-model}
		H = \sum_{a=1}^{m}\sum_{1\le i,j\le N}(h_{ij} - \mu \delta_{ij})\psi_{i,a}^{+}\psi_{j,a}^{-}\ed 
	\end{align}
	Here, $\mu$ is the chemical potential, $i,j = 1,2,\ldots,N$ label the sites, and $a = 1,2,\ldots,m$ is the flavor index.  We always impose the unitary condition in Eq.~\eqref{eq:unitary-condition}, ensuring that $\psi_{i,a}^{+} = \psi_{i,a}^{\dagger}$ throughout the paper. The term $h_{ij}$ is a random Hermitian matrix drawn from GUE with $h_{ii}\sim \mathcal{N}(0,1/N)$, $\text{Re}(h_{ij})\sim \mathcal{N}(0,{1\over 2N})$ and $\text{Im}(h_{ij})\sim \mathcal{N}(0,{1\over 2N})$. Here, $\mathcal{N}(u,\sigma^2)$ denotes a normal distribution with mean $u$ and variance $\sigma^2$. The model can be transformed into a free model via the substitution $\psi_{i,a}^- \to U_{ij}\widetilde{\psi}_{j,a}^-$, where $U_{ij}$ is the matrix that diagonalizes $h_{ij}$. The new operators satisfy the same commutation relations, so we will not distinguish them. This yields
	\begin{align}
		H=\sum_{i=1}^N (\varepsilon_i-\mu)n_i, n_i=\sum_{a=1}^m \psi_{i,a}^{+}\psi_{i,a}^{-}
	\end{align} 
	where $\varepsilon_j\in \mathbb{R}$ are the eigenvalues of the coupling matrix $h_{ij}$. 
	So, formally, we can generalize the model to any case with an arbitrary adjoint probability $P(\varepsilon_1,\varepsilon_2,\ldots, \varepsilon_N)$. In this paper, we will focus on CUE while other cases (GSE,GOE,CSE,COE) are also discussed. 
	Unlike Gaussian ensembles, the circular ensembles have complex eigenvalues $\{e^{i\theta_j}\},\theta_j\in (-\pi,\pi]$.\footnote{In this paper, we occasionally use the notation $\ii\equiv \sqrt{-1}$ to prevent confusion.} There are two choices
	\begin{align}
		(i): \varepsilon_j=e^{i\theta_j},\text{or}\ (ii):\varepsilon_j = \theta_j\ed
	\end{align}
	The first one is nothing but taking $h_{ij}$ as the elements of the circular ensemble, but it will lead to non-Hermitian Hamiltonian.  Non-Hermitian Hamiltonians make sense as an effective description of open quantum systems, and non-Hermitian SYK models have been studied in \cite{Garcia-Garcia:2021rle,Garcia-Garcia:2022rtg}. It is suggested that complex spacing ratios serve as a quantum chaos diagnostic for non-Hermitian Hamiltonian and general open quantum systems. We find the SFF of the first choice is a constant, so in this paper, we will focus on the second one.  At first glance, it lacks a physical meaning and we need to interpret $\theta_j$ as quasi-energy, but it preserves the Hermitian Hamiltonian. Moreover, in the thermodynamic limit $N\to \infty$, the correlation and cluster functions of the GUE and CUE are known to
	agree asymptotically. Thus, studying the SFF with circular ensembles can serve as a benchmark for the conclusions drawn from Gaussian ensembles.
	
	As shown in \cite{Li:2025kpz}, the grand canonical partition function and SFF of the model  \textbf{only} depend on the dimension of the $n$-particle Hilbert space (or the degeneracy of the $n$-th level), denoted by $d_n (n = 0,1,2,\ldots)$. The partition function and SFF are defined as 
	\begin{align}
		Z_{\beta}=\prod_{j=1}^N \sum_{n=0}^L d_n e^{-\beta n (\varepsilon_j-\mu)},\quad K_t= {1\over D^{2N}}\prod_{j=1}^N \abs{\sum_{n=0}^L d_n e^{-i t n (\varepsilon_j-\mu)}}^2
	\end{align}
	where $D\equiv \sum_n d_n$ is the total dimension of the ``local" Hilbert space. We denote their ensemble average by $\mathcal{Z}_\beta= \langle Z_\beta \rangle, \mathcal{K}_t=\langle K_t \rangle$. 
	The value of $d_n$ is determined by the choice of the $R$-matrix. This information is encoded in the single-mode partition function
	\begin{align}
		z_R(x)\equiv \sum_{n=0}^\infty d_n x^n\ed
	\end{align}
	For ordinary fermions and bosons with $m$ flavors, the single-mode partition functions are given by  
	\begin{align}  
		z_{\text{Fermion}}(x) = (1 + x)^{m}, \quad  
		z_{\text{Boson}}(x) = (1 - x)^{-m}.  
	\end{align}  
	Here we adopt a simplified terminology: \textit{$R$-para-fermions} refer to $R$-para-particles with a finite polynomial partition function $z_R(x) = \sum_{n=1}^L d_n x^n$ (where $L$ is finite), while \textit{$R$-para-bosons} describe $R$-para-particles with a fractional $z_R(x)$.
	In this work, we focus on the study of SFF for \textit{$R$-para-fermionic} SYK$_2$ (Example A and B in Table \ref{table:six-cases}), while the partition function can be calculated in a similar manner. Moreover, we mainly discuss the behavior of SFF in the time regime $t\ge 1$.  
	\paragraph{Main results}
	In this paper, we primarily evaluate the averaged spectral form factor (SFF) of the $R$-PSYK$_2$ model with different matrix ensemble couplings. Assuming an analogy between Gaussian and Circular ensembles, we first derive an exact analytical match between the GUE and CUE results for Example A. Additionally, we numerically verify a consistent agreement between GUE and CUE for Example B. Based on these findings, we propose a relationship in the time regime $1 < t \ll N$ in the large $N$ limit, expressed as
	\begin{align}  
		\mathcal{K}_{\text{GUE}}(2t) \approx \mathcal{K}_{\text{CUE}}(t).  
	\end{align}  
	In \cite{Michael_2024}, the authors scaled time by relevant single-particle Heisenberg time, namely $\mathcal{K}_{\text{GUE}}(\pi/\sqrt{2} t) \approx \mathcal{K}_{\text{CUE}}(t)$ for any time $t>1$. In this paper, we choose the scaling factor of 2 to match the time at which the PSYK SFF of GUE and CUE reaches the plateau. It can be observed that with this choice, the analytical results for GUE and CUE correspond precisely. We further examine the similarity between GSE and CSE numerically, observing a slope transition in the SFF behavior for CSE from the numerical data. 
	\paragraph{Structure of the paper}
	The paper is structured as follows. Section \ref{sec:GaussinE} delves into the Gaussian ensembles, specifically GUE, GSE, and GOE. The SFF for these ensembles is analyzed, with detailed derivations for GUE and GSE, while the challenges for GOE are noted. Section \ref{sec:CircularE} explores the circular ensembles, including CUE, CSE, and COE. The SFF for CUE is derived analytically, and numerical comparisons with GUE are provided. The section also discusses the behavior of SFF for CSE. For COE, we outline the calculation method but omit numerical results. Section \ref{sec:discussion} concludes the paper, summarizing the findings and suggesting directions for future research. 
	\section{Gaussian Ensembles}
	\label{sec:GaussinE}
	We first study the three Gaussian ensembles, namely GUE, GSE and GOE. For GUE coupling, this corresponds to the case studied in \cite{Li:2025kpz}. While the standard complex SYK$_2$ model can be analyzed via path integrals and saddle-point methods to compute thermodynamics, two-point functions, and the SFF, we omit this approach here due to its limited generalizability to $R$-PSYK$_2$. Instead, we focus on random matrix techniques.
	
	Generally, $N\times N$ ($2N\times 2N$ for GSE, due to symplectic symmetry) Hermitian matrices from the three Gaussian ensembles have the following joint probability distribution for their eigenvalues $\varepsilon_j\in \mathbb{R}$
	\begin{align}\label{eq:Pb}
		P_{\mathsf{b}}(\varepsilon_{1},\ldots,\varepsilon_{N})=C_{N,\mathsf{b}}\exp\left(-\frac{1}{2}N\sum_{j=1}^{N}\varepsilon_{j}^{2}\right)\prod_{j<k}|\varepsilon_{j}-\varepsilon_{k}|^{\mathsf{b}}
	\end{align}
	where $\mathsf{b} = 1, 2, 4$ corresponds to GOE (real symmetric), GUE (complex Hermitian), and GSE (quaternion Hermitian),  respectively.  We use $\mathsf{b}$ to distinguish from the inverse temperature $\beta$. For a $2N\times 2N$ dimensional matrix from GSE, the spectrum is two-fold degenerate, so we represent its joint probability distribution using $N$ parameters. Note that Eq.~\eqref{eq:Pb} differs from the conventional form found in standard references \cite{mehta2004} due to our specific scaling, consistent with what we implemented for GUE in Eq.~\eqref{eq:original-model}. This formulation agrees with \cite{Liao_2020}. The normalization factor is given by
	\begin{align}
		C_{N,\mathsf{b}}=N^{\frac{N}{2}+\frac{\mathsf{b} N(N-1)}{4}}\left(2\pi\right)^{-N/2}\prod_{j=1}^{N}\frac{\Gamma\left(1+\mathsf{b}/2\right)}{\Gamma\left(1+\mathsf{b}j/2\right)}\co
	\end{align}
	ensures unit total probability.
	For later convenience, we introduce the notation
	\begin{align}
		\Delta_{N}^{\mathsf{b}}(\varepsilon)=\prod_{j<k}|\varepsilon_{j}-\varepsilon_{k}|^{\mathsf{b}}\ed
	\end{align}
	Since both the partition function and SFF can be expressed as products $G(\{\varepsilon\})=\prod_{j=1}^N \mathcal{G}(\varepsilon_j)$, our goal is to evaluate
	\begin{align}
		\langle G \rangle =\int\prod_{j=1}^{N}d\varepsilon_{j}\mathcal{G}\left(\varepsilon_{j}\right)P_{\mathsf{b}}(\varepsilon_{1},\ldots,\varepsilon_{N})=\int\Delta_{N}^{\mathsf{b}}(\varepsilon)\prod_{j=1}^{N}d\varepsilon_{j}\mathcal{G}\left(\varepsilon_{j}\right)\left(C_{N,b}\right)^{1/N}e^{-\frac{1}{2}N\varepsilon_{j}^{2}}\ed
	\end{align}
	Performing the substitution $\varepsilon_{j}\to\sqrt{\frac{2}{N}}x_{j}$ and defining
	\begin{align}
		w_{\mathsf{b}}(x)=\left(\frac{2}{N}\right)^{\mathsf{b}\frac{(N-1)}{4}+1/2}\left(C_{N,\mathsf{b}}\right)^{1/N}e^{-x^2}
		\equiv W_\mathsf{b} e^{-x^2}\co
	\end{align}
	we obtain the final expression
	\begin{align}\label{eq:Gauss-G}
		\langle G \rangle &=\frac{2^{\frac{\mathsf{b}N(N-1)}{4}}}{\pi^{N/2}}\prod_{j=1}^{N}\frac{\Gamma\left(1+\mathsf{b}/2\right)}{\Gamma\left(1+\mathsf{b}j/2\right)}\int\Delta_{N}^{\mathsf{b}}(x)\prod_{j=1}^{N}dx_{j}\mathcal{G}\left(\sqrt{\frac{2}{N}}x_{j}\right)e^{-x_{j}^{2}}\ed
	\end{align}
	The crucial property enabling exact analysis is that the SFF (and partition function) can be expressed as a product of the \textbf{same} function evaluated at different single variables $x_j$. There are two approaches to evaluate the integral in Eq.~\eqref{eq:Gauss-G}. 
	
	The first method involves expanding $\Delta_{N}^{\mathsf{b}}(x)$ as a sum of products of single-variable functions, which factorizes the integral and ultimately requires computing the determinant of an $N$-dimensional ($2N$-dim for GSE) matrix $\boldsymbol{\alpha}$. This approach suggests choosing an appropriate expansion for $\Delta_{N}^{\mathsf{b}}(x)$ such that $\boldsymbol{\alpha}$  becomes a truncated version of some operator $\hat{O}$ in a simple quantum system (harmonic oscillator for GUE and GSE cases). However, computing $\det\boldsymbol{\alpha}$ is generally quite complex. As demonstrated in \cite{Liao_2020}, this requires evaluating traces of matrices whose elements are expressed in terms of Laguerre polynomials. In the large $N$ limit, obtaining analytical expressions becomes particularly challenging, especially for the time regime $t \ll N$.
	
	\paragraph{Coherent state approach} The main challenge is determining the eigenvalues of $\boldsymbol{\alpha}$. For our model, we find it more tractable to diagonalize the infinite-dimensional operator $\hat{O}$ rather than its finite truncation $\boldsymbol{\alpha}$. In the large $N$ limit, we can approximate the solution by first diagonalizing $\hat{O}$ in an appropriate basis and then implementing a suitable eigenvalue cutoff. For the GUE and GSE cases discussed here, we select coherent states as our basis, leading to what we term the coherent state approach.
	
	The validity of this approach for partition function calculations (at least at finite temperature) was established in \cite{Li:2025kpz}. Furthermore, this method proves effective in studying the SFF in the early-time regime $0 < t \lesssim \mathcal{O}(1)$ and correctly predicts the plateau height for $t \gg N$. Its primary limitation is the inability to capture the ramp behavior in the intermediate region $1 \ll t \ll N$, which is precisely our region of interest. Consequently, while we derive the relevant formulas here, we omit detailed calculations.
	\begin{table}[ht]\label{table:coeff}
		\centering
		\vspace{0.5cm}
		\begin{tabular}{l|l|l} 
			\toprule
			& \textbf{GUE,GOE,CUE,COE} & \textbf{GSE,CSE} \\
			\midrule
			Exp.A & $\{\widetilde{g}_k\}_{k=0}^1=\{\frac{m^2+1}{(m+1)^2},\frac{2 m}{(m+1)^2}\}$ & $\{\widetilde{g}_k\}_{k=0}^2=\{\frac{m^{4}+4m^{2}+1}{(m+1)^{4}},\frac{4m^{3}+4m}{(m+1)^{4}},\frac{2m^{2}}{(m+1)^{4}}\}$\\
			Exp.B & $\{\widetilde{g}_k\}_{k=0}^2=\{\frac{m^2+2}{(m+2)^2},\frac{4 m}{(m+2)^2},\frac{2}{(m+2)^2}\}$ & $\{\widetilde{g}_k\}_{k=0}^3=\{\frac{m^4+12 m^2+6}{(m+2)^4},\frac{8 m^3+24 m}{(m+2)^4},\frac{12 m^2+8}{(m+2)^4},\frac{8
				m}{(m+2)^4},\frac{2}{(m+2)^4}\}$ \\
			\bottomrule
		\end{tabular}
		\caption{The normalized coefficients used in calculating SFF. }
	\end{table}

	\paragraph{Cluster function approach} The second approach utilizes the symmetry of the joint probability distribution with respect to its arguments, known as the cluster function approach as presented in \cite{Michael_2024}. We briefly review this method here.
	For a product $G=\prod_{j=1}^N (1+F(\varepsilon_j,t))$ with arbitrary function $F(\varepsilon,t)$, its expectation reads
	\begin{equation}\label{eq:G-rn-exp}
		\begin{aligned}
			\langle G \rangle  
			&= \int\prod_{i=1}^{N}d\varepsilon_{i}P_{N}(\varepsilon_{1},\ldots,\varepsilon_{N})\prod_{j=1}^{N}\left[1+F\left(\varepsilon_{j},t\right)\right] \\
			&= 1+\sum_{n=1}^{N}\frac{1}{n!}\int\mathsf{R}_{n}(\varepsilon_{1},\ldots,\varepsilon_{n})\prod_{j=1}^{n}F\left(\varepsilon_{j},t\right)d\varepsilon_{j} 
			\equiv 1+\sum_{n=1}^{N}\frac{1}{n!}\mathsf{r}_n\ed
		\end{aligned}	
	\end{equation}
	The $n$-point single-particle energy level correlation function $\mathsf{R}_{n}(\varepsilon_{1},\ldots,\varepsilon_{n})$ is defined as
	\begin{align}
		\mathsf{R}_{n}(\varepsilon_{1},\ldots,\varepsilon_{n}) = \frac{N!}{(N-n)!}\int d\varepsilon_{n+1}\ldots d\varepsilon_{N}P_{N}(\varepsilon_{1},\ldots,\varepsilon_{N}).
	\end{align}
	In the large $N$ limit, the correlation function is determined by the determinant of the kernel $\mathsf{K}(\varepsilon_i,\varepsilon_j)$
	\begin{align}
		\mathsf{R}_n(\varepsilon_{1},\varepsilon_{2},\ldots,\varepsilon_{n})=\det\left[\mathsf{K}(\varepsilon_{i},\varepsilon_{j})\right]_{i,j=1,\ldots,n}
	\end{align} 
	where (for GUE)
	\begin{align}
		\mathsf{K}(\varepsilon_i,\varepsilon_j) = 
		\begin{cases} 
			\frac{N}{2\pi}\sqrt{4-\varepsilon_i^2}\,\Theta(2-|\varepsilon_i|), & i=j, \\[10pt]
			\frac{N}{\pi}\frac{\sin\left[N(\varepsilon_i-\varepsilon_j)\right]}{N(\varepsilon_i-\varepsilon_j)}, & i\neq j.
		\end{cases}
	\end{align}
	Here $\Theta(x)$ is the Heaviside theta function. One can introduce the $n$-point cluster function 
	\begin{align}
		\mathsf{T}_n(\varepsilon_1, \dots, \varepsilon_n) = \sum_{\mathcal{P}(n)} \mathsf{K}(\varepsilon_1, \varepsilon_2) \cdots \mathsf{K}(\varepsilon_{n-1}, \varepsilon_n) \mathsf{K}(\varepsilon_n, \varepsilon_1),
	\end{align}
	where the sum runs over $(n-1)!$ cyclic permutations $\mathcal{P}(n)$ of indices $\{1,2,\ldots, n\}$. This leads to the compact large $N$ expression
	\begin{align}
		\langle G \rangle \approx \exp\left[\sum_{n=1}^{N}\frac{(-1)^{n-1}}{n!}\mathsf{t}_{n}\right],~\mathsf{t}_{n} = \int d\varepsilon_{1}\ldots d\varepsilon_{n}\,\mathsf{T}_{n}(\varepsilon_{1},\dots,\varepsilon_{n})\prod_{i=1}^{n}F(\varepsilon_{i},t)\ed
	\end{align}
	For the SFF calculation, we consider the expansion
	\begin{align}
		\prod_{j=1}^{N} \abs{\sum_{n_{j}=0}^{L} d_{n_j}e^{-\ii \theta_{j} n_{j}t}}^2 = D^{2N}\prod_{j=1}^N(1+F(\varepsilon_j,t))
	\end{align}
	where we define $\abs{\sum_{n=0}^{L} d_{n}e^{-\ii \varepsilon n t}}^2 \equiv D^2(1+F(\varepsilon,t)).$
	This yields the expansion
	\begin{equation}
		\begin{aligned}
			\prod_{j=1}^{n}F(\varepsilon_{j},t) &= \prod_{j=1}^{n}\left[\widetilde{g}_{0}-1+\sum_{k=1}^{L}\widetilde{g}_{k}\cos\left(kt\left(\varepsilon_{j}-\mu\right)\right)\right] \\
			&= \frac{1}{2^{n}}\sum_{\{\zeta\}=-L}^{L}\mathsf{g}_{\{\zeta\}}\exp\left(it\sum_{j=1}^{n}\zeta_{j}\left(\varepsilon_{j}-\mu\right)\right)
		\end{aligned}
	\end{equation}
	where we have used the shorthand $\mathsf{g}_{\{\zeta\}}=\prod_{j=1}^n \mathsf{g}_{\zeta_j}$ with coefficients defined by
	\begin{align}
		\mathsf{g}_{0}=2\left(\widetilde{g}_{0}-1\right),\quad \mathsf{g}_{k>0}=\mathsf{g}_{-k}=\widetilde{g}_{k}\ed
	\end{align}
	For reference in subsequent calculations, we summarize the expressions for $\widetilde{g}$ coefficients used throughout this paper in Table~\ref{table:coeff}.

	\subsection{GUE}
	We first review the approach to GUE and note the equality
	\begin{align}\label{eq:Delta-2-exp}
		\Delta_N^2(x) = \sum_{(i)} \sum_{(j)} \sigma(i) \sigma(j) C_{i_1}(x_1) \ldots C_{i_N}(x_N) \overline{C}_{j_1}(x_1) \ldots \overline{C}_{j_N}(x_N),
	\end{align}
	where  $C_i(x)$ is monic polynomial
	of precise degree $i$ and $\overline{C}_i$ is its conjugate.  Recall that a polynomial is called monic when the coefficient of the
	highest power, i.e. $x^i$, is one. And \( \sigma(i) \) is the sign of the permutation \( (i) = \binom{0 \ldots N-1}{i_1 \ldots i_N} \), the sum \( (i) \) is over all the \( N! \) permutations \( (i) \), and similarly for the permutations \( (j) \). We can choose $C_n(x)=2^{-n}H_{n}(x)$,~$H_{k}(x)=2^{k/2}\sqrt{k!}\pi^{1/4}\mathsf{H}_{k}(x)$, where $\mathsf{H}_{k}(x)$ are the normalized Hermite polynomials. Using Eq.~\eref{eq:Gauss-G}, we finally have 
	\begin{equation}
		\begin{aligned}
			\langle G \rangle 
			&={1\over N!}\int \sum_{(i)(j)}  \sigma(i) \sigma(j) \mathsf{H}_{i_1}(x_1) \ldots \mathsf{H}_{i_N}(x_N) \mathsf{H}_{j_1}(x_1) \ldots \mathsf{H}_{j_N}(x_N)\prod_{l=1}^{N}dx_{l}\mathcal{G}\left(\sqrt{\frac{2}{N}}x_{l}\right)e^{-x_{l}^{2}}\\
			&=\det \mathsf{M}
		\end{aligned}
	\end{equation}
	where 
	\begin{align}\label{eq:coherent-M}
		\mathsf{M}_{ij} = \int dx\, e^{-x^{2}}\mathsf{H}_{i-1}(x)\mathsf{H}_{j-1}(x)\mathcal{G}\left(x\sqrt{\frac{2}{N}}\right)
		= \langle i-1|\mathcal{G}\left(\hat{x}\sqrt{\frac{2}{N}}\right)|j-1\rangle.
	\end{align}
	Noting that $e^{-x^2/2}\mathsf{H}_k(x)$ is the $k$-th wavefunction of a quantum harmonic oscillator with $m = \omega = 1$, (here $k=0$ corresponding to the ground state).  We can treat $x$ as an operator. We evaluate the large $N$ limit as
	\begin{align}\label{eq:G_trace}
		\frac{1}{N}\log\langle G \rangle = \frac{1}{N}\log\det\mathsf{M} \approx \frac{1}{N}\sum_{i=1}^{N}\log\lambda_{i}
	\end{align}
	where $\lambda_i$ is the $i$-th eigenvalue of the operator $\hat{\mathcal{G}}\equiv\mathcal{G}\left(\hat{x}\sqrt{\frac{2}{N}}\right)$. Note that $\mathsf{M}$ is an $N$-dimensional matrix obtained by truncating the infinite-dimensional matrix $\hat{\mathcal{G}}$, so its eigenvalues are not exactly the same. However, we expect the approximation in Eq.~\eqref{eq:G_trace} to hold for large $N$. The trace in Eq.~\eqref{eq:G_trace} can be interpreted as a truncated trace in the harmonic oscillator Hilbert space
	\begin{align}
		\text{tr}_N(\bullet) \equiv \text{tr}(\mathbb{I}_N\bullet \mathbb{I}_N) = \sum_{k=0}^{N-1}\langle k|\bullet|k\rangle \Rightarrow \frac{1}{N}\log\langle G \rangle \approx \frac{1}{N}\text{tr}_N\log(\hat{\mathcal{G}})
	\end{align}
	where $\mathbb{I}_N = \sum_{i=0}^{N-1}|i\rangle\langle i|$ is the projector onto the first $N$ states. For the coherent state basis, we restrict the trace to states with $|\alpha| \leq R$ and take the following replacements
	\begin{align}\label{eq:trN-to-trR}
		N &= \text{tr}_R\mathbb{I} = R^{2}, \\
		\text{tr}_N A &\to \text{tr}_R(A) = \int_{|\alpha|<R}\frac{d^2\alpha}{\pi}A(\alpha,\alpha^{*}), \\
		\hat{x}\sqrt{\frac{2}{N}} &\to \frac{1}{R}(\hat{a} + \hat{a}^{\dagger})\ed
	\end{align}
	Using the representation $\hat{a}^\dagger \equiv \frac{1}{\sqrt{2}}\left(\hat{x}-i\hat{p}\right)= \alpha^*-\partial_\alpha, \hat{a}\equiv \frac{1}{\sqrt{2}}\left(\hat{x}-i\hat{p}\right)=\alpha$ in coherent state basis, we finally have the approximation for large $N$
	\begin{equation}\label{eq:Coherence-appro}
		\begin{aligned}
			\frac{1}{N}\log\langle G\rangle \approx \frac{1}{R^{2}}\text{tr}_{R}\log\left[\mathcal{G}\left(\frac{a + a^{\dagger}}{R}\right)\right] 
			\approx \int_{0}^{1}\int_{0}^{2\pi}\frac{u\,du\,d\theta}{\pi}\log\left[\mathcal{G}(2u\cos\theta)\right]\ed
		\end{aligned}
	\end{equation}
	As demonstrated in \cite{Li:2025kpz}, the coherent state approach is well-suited for calculating the partition function at high temperatures and SFF at early times. While this method correctly reproduces the late-time plateau of the SFF, it fails to capture the ramp behavior in the intermediate time regime. To address this limitation, we instead employ the cluster function approach.
	For GUE-coupled systems, the cluster function method has been extensively studied in earlier works \cite{Liao_2020, Li:2025kpz}. Since the technical details are already well-documented, we focus our discussion here on presenting the key results. The SFF in $t_p \equiv (N/C_0)^{2/5}\ll t \ll N$ for $R$-PSYK$_2$ model with GUE couplings has the expression 
	\begin{equation}
		\mathcal{K}_{t_p\ll t\ll N} =D^N \widetilde{g}_0^N\exp\left[NB_{0} + C_0t\right]
	\end{equation}
	with coefficients are defined as
	\begin{equation}
		\begin{aligned}\label{eq:BpCp-for-num}
			B_{0} = \sum_{n=2}^{N}\frac{(-1)^{n-1}}{n2^n}\sum_{\substack{\sum\zeta_{i}=0}}\mathsf{g}(\{\zeta_{i}\}), ~
			C_{0} = \sum_{n=2}^{N}\frac{(-1)^{n}}{n2^{n+1}}\sum_{\substack{\sum\zeta_{i}=0}}\mathsf{g}(\{\zeta_{i}\})s(\{\zeta_{i}\})\ed 
		\end{aligned}
	\end{equation}
	Here $\zeta_i$ takes values in $\{-L,\ldots,0,\ldots,L\}$ and 
	$s(\{\zeta_{i}\})$ is defined as
	\begin{equation}\label{eq:s-factor}
		s(\{\zeta_i\}) = \max\left\{0, \sum_{i=1}^j \zeta_i\right\}_{j=1}^{n-1} - \min\left\{0, \sum_{i=1}^j \zeta_i\right\}_{j=1}^{n-1}.
	\end{equation} 
	Consider paths generated by displacements $\{\zeta_i\}$ where $x_j = \sum_{i=1}^j \zeta_i$ with $x_0 = 0$. Each path has displacements $\zeta_{j+1} \in \{0, \pm1, \pm2, \ldots, \pm L\}$ weighted by $\mathsf{g}_{|\zeta_{j+1}|}$. The factor $s(\{\zeta_i\}) = \max(\{x\}) - \min(\{x\})$ accounts for the vertical extent of each path. One can numerically calculate $B_0,C_0$ using the transfer matrix approach displayed in \cite{Li:2025kpz}. The transfer matrix $T^{\le k}$, with dimension $(k+\lfloor nL/2 \rfloor +1)\times (k+\lfloor nL/2 \rfloor +1)$, is a symmetric Toeplitz matrix with bandwidth $L$
	\begin{align}
		T^{\le k} = 
		\begin{bmatrix} 
			\mathsf{g}_0 & \mathsf{g}_1 & \cdots & \mathsf{g}_L & 0 & \cdots \\
			\mathsf{g}_1 & \mathsf{g}_0 & \mathsf{g}_1 & \ddots & \ddots & \\
			\vdots & \ddots & \ddots & \ddots & \ddots & \mathsf{g}_L \\
			\mathsf{g}_L & \ddots & \ddots & \ddots & \ddots & 0 \\
			0 & \ddots & \ddots & \ddots & \ddots & \vdots \\
			\vdots & & \mathsf{g}_L & \cdots & \mathsf{g}_1 & \mathsf{g}_0
		\end{bmatrix}.
	\end{align}
	The final expressions for the coefficients are
	\begin{align}
		C_{0} &= \sum_{n=2}^{N}\frac{(-1)^{n}}{n2^{n}}\left[(\lfloor nL/2\rfloor+1)\bra{0}T^n\ket{0} - \sum_{k=0}^{\lfloor nL/2\rfloor}\bra{0}\left(T^{\le k}\right)^n\ket{0}\right], \\
		B_{0} &= \sum_{n=2}^{N}\frac{(-1)^{n-1}}{n2^n}\bra{0}T^n\ket{0}\co
	\end{align}
	where $T \equiv T^{\le \lfloor nL/2\rfloor}$. The state $\ket{x}$ is a unit vector with value 1 at position $(x + \lfloor nL/2 \rfloor + 1)$ and zero elsewhere.
	
	For Example A with $z_R(x) = 1 + mx$, we derive the exact asymptotic form of $C_0$
	\begin{align}
		C_{0} = 
		\begin{cases}
			\dfrac{1}{2}\ln\left(\dfrac{m^{2}}{m^{2}-1}\right), & m > 1, \\[10pt]
			\dfrac{1}{2}\left(\ln\dfrac{N}{8} + \gamma_{E}\right), & m = 1,
		\end{cases}
	\end{align}
	where $\gamma_E$ denotes the Euler-Mascheroni constant.
	For Example B with $z_R(x) = 1 + mx + x^2$, analytical solutions remain elusive. Numerical simulations show that: cases with $m=1,2$ require fitting using data in the range $10 < N \leq 100$ while all other cases converge rapidly. 
	The corresponding results are presented in Table~\ref{table:C0-GUE}.

	\begin{table}[ht]
		\centering
		\vspace{0.5cm}
		\begin{tabular}{l|llll} 
			\toprule
			$C_0$& $m=1$ & $m=2$ & $m=3$&$m=4$ \\
			\midrule
			Example A & $	\frac{1}{4}\left(\ln\frac{N}{8}+\gamma_{E}\right)$ & ${1\over 2}\ln{4\over 3}$ &${1\over 2}\ln{9\over 8}$ &${1\over2 }\ln {16\over 15}$\\
			Example B & $0.528818 N^{0.25}-0.672036 $ & $0.672053 N^{0.2}-0.813986 $&$0.314624$ &$0.149009$ \\
			\bottomrule
		\end{tabular}
		\caption{The results of $C_0$ for two examples considered in this paper, where $\gamma_E \approx  0.577$ is the Euler-Mascheroni constant. }
		\label{table:C0-GUE}
	\end{table}
	To verify our theoretical results, we can perform numerical simulations directly according to the model definition by taking multiple sample averages. Alternatively, we may compute the matrix elements following Eq.~\eref{eq:coherent-M} and then evaluate the determinant. For our current purposes, we adopt the determinant approach and, for simplicity, choose the most elementary polynomial form: $C_j(x)=x^j$. 
	From Eq.~\eref{eq:Delta-2-exp}, we have 
	\begin{align}\label{eq:GUE-SFF-num}
		\mathcal{K}(t)=D^{2N}N!\det(\boldsymbol{\alpha})
	\end{align}
	with
	\begin{align}
		\mathsf{\alpha}_{ij}&=\int\mathcal{G}(\sqrt{\frac{2}{N}}x)w_{2}(x)x^{i+j}dx=\frac{W_{2}}{2}\sum_{\xi=\pm1}\sum_{k=1}^{L}\widetilde{g}_{k}I_{i+j}(\xi kt\sqrt{\frac{2}{N}})e^{-i\xi kt\mu}\ed
	\end{align}
	Here we have defined
	\begin{equation}
		\begin{aligned}
			I_{n\ge 0}(u)&\equiv \int dxe^{-x^{2}}x^{n}e^{iux}=\delta_{\text{even}}(n)\Gamma\left(\frac{n+1}{2}\right)\,_{1}F_{1}\left(\frac{n+1}{2};\frac{1}{2};-\frac{u^{2}}{4}\right)\\
			&\quad +i\delta_{\text{odd}}(n)u\Gamma\left(\frac{n}{2}+1\right)\,_{1}F_{1}\left(\frac{n}{2}+1;\frac{3}{2};-\frac{u^{2}}{4}\right)\ed 
		\end{aligned}
	\end{equation}
	Here $\,_{1}F_{1}(a;b;z)$ is the Kummer confluent hypergeometric function, and 
	\begin{align}
		\delta_{\text{even}}(n)=\begin{cases}
			1, & n\ \text{even}\\
			0, & n\ \text{odd}
		\end{cases},~\delta_{\text{odd}}(n)=\begin{cases}
			0, & n\ \text{even}\\
			1, & n\ \text{odd}
		\end{cases}\ed
	\end{align}
	\begin{figure}[ht]
		\begin{center}
			\includegraphics[width=0.45\textwidth]{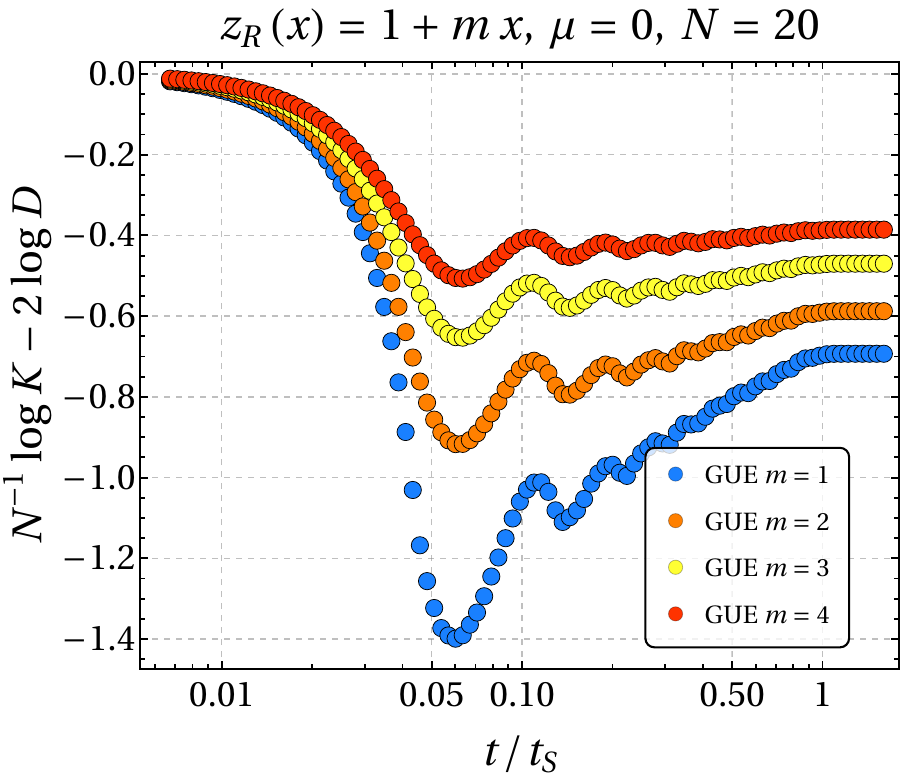}
			\includegraphics[width=0.45\textwidth]{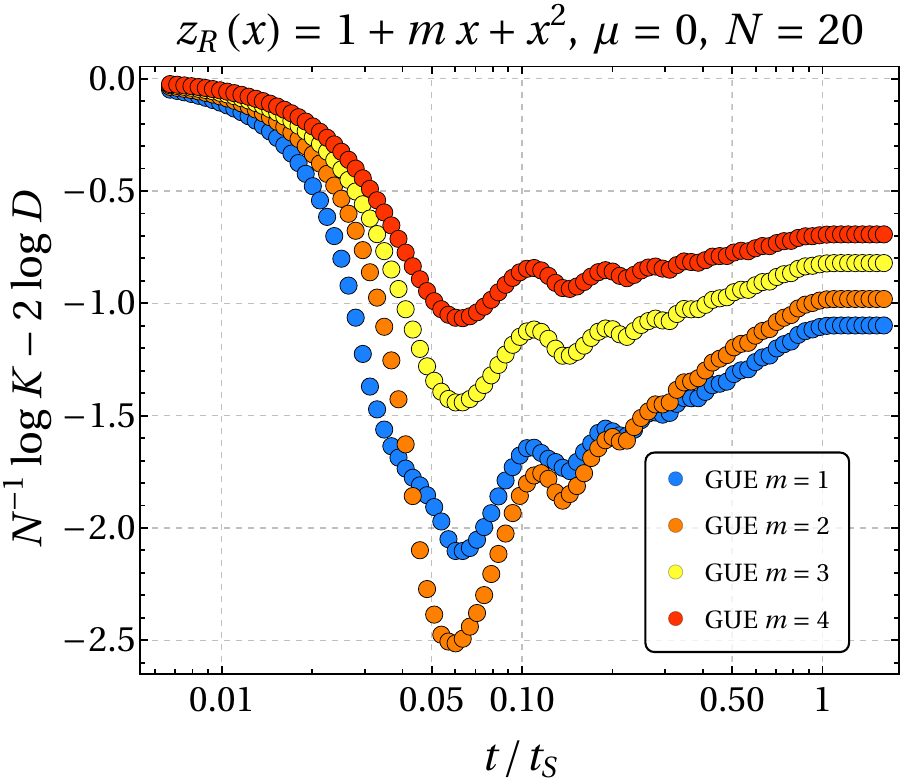}
			\caption{Numerical results for the SFF of two {GUE} instances are presented, where we set the scaling time $t_S = 2N$. In the large $N$ limit, the theoretic prediction implies that $\frac{1}{N} \log \mathcal{K}$ depends linearly on $t$ within the time regime $t_p \ll t \ll N$. For finite $N$, however, oscillations obscure the numerical results. One may average over multiple $N$ values to reveal a distinct linear growth regime. In this work, we instead leverage the similarity between {GUE} and {CUE} to perform an exact analytical verification.    }
			\label{fig:GUE-m}
		\end{center}
	\end{figure}
	
	Although our theoretical analysis predicts the existence of an \textbf{exponential} ramp in the SFF, with a growth rate that transitions from infinity to a finite value as $m$ varies (see Table~\ref{table:C0-GUE}), direct numerical verification presents significant challenges. As demonstrated in \cite{Li:2025kpz}, conventional random matrix simulations fail to clearly detect this ramp behavior, making precise validation of our theoretical predictions difficult.
	
	Rather than relying on direct simulation, we instead employ Eq.~\eref{eq:GUE-SFF-num}, which provides equivalent results to numerical simulations in the large $N$ limit. Figure~\ref{fig:GUE-m} shows two representative cases, where we observe that the divergent $C_0$ case exhibits significantly stronger SFF growth. This only offers qualitative confirmation of our theoretical framework.
	
	The difficulty in numerically observing the ramp stems from two key factors:
	\begin{itemize}
		\item The SFF contains oscillatory terms that mask the underlying ramp behavior.
		\item The ramp only appears in the time window $(N/C_0)^{2/5} \ll t \ll N$, requiring extremely large $N$ to create a discernible temporal regime.
	\end{itemize}
	Practical limitations in machine precision make simulations with sufficiently large $N$ values computationally prohibitive, explaining why the ramp remains elusive in numerical approaches.
	\subsection{GSE}
	Next, we will address the case of GSE. Following the approach used for GUE in the previous section, the key step is to expand the determinant factor as a product of univariate polynomials. We find that for GSE, Hermite polynomials can still be employed, and the computation of the SFF reduces to evaluating the determinant of a truncated quantum harmonic oscillator operator, though the operator here differs from that in GUE. Similarly, we can also discuss the coherent state method, and we observe that the final result resembles that of GUE. Therefore, their SFFs exhibit similar early-time behavior. However, this is not the main focus of this paper, so we do not plan to conduct further numerical verification on this matter. Likewise, this method can also be applied to compute the ensemble-averaged partition function, though we will not elaborate on it here.
	
	For GSE, $\mathsf{b}=4$, and the dimension of matrix $h_{ij}$ is $2N$. Namely, the model is defined as 
	\begin{align}
		H = \sum_{a=1}^{m}\sum_{1\le i,j\le 2N}(h_{ij} - \mu \delta_{ij})\psi_{i,a}^{+}\psi_{j,a}^{-}\ed
	\end{align} 
	The basic strategy is the identity
	\begin{align}
		\Delta_{N}^{4}(x)=\sum_{(i)}\sigma(i)\det\begin{bmatrix}Q_{i_{1}}(x_{1}) & Q_{i_{1}}^{\prime}(x_{1})\\
			Q_{i_{2}}(x_{1}) & Q_{i_{2}}^{\prime}(x_{1})
		\end{bmatrix}\cdots\det\begin{bmatrix}Q_{i_{2N-1}}(x_{N}) & Q_{i_{2N-1}}^{\prime}(x_{N})\\
			Q_{i_{2N}}(x_{N}) & Q_{i_{2N}}^{\prime}(x_{N})
		\end{bmatrix}
	\end{align}
	where \(\sigma(i)\) is the sign of the permutation \((i) = \binom{0 \ldots 2N-1}{i_1 \ldots i_{2N}}\), the sum \((i)\) is over all of those \((2N)!\) permutations \((i)\) which satisfy the restrictions \(i_1 < i_2, \ldots, i_{2N-1} < i_{2N}\). Notice that each variable occurs in only one \(2 \times 2\) sub-determinant. So recalling the definition of Pfaffian, one find the result is
	\begin{equation}
		\langle G \rangle  = N! \, \operatorname{pf}[\alpha_{ij}]_{i,j=0,\ldots,2N-1} =N! \sqrt{\det\boldsymbol{\alpha}}
	\end{equation}
	where 
	\begin{equation}
		\alpha_{ij} = \int \mathcal{G}(\sqrt{\frac{2}{N}}x) w_{4}(x) \left(
		Q_i(x) Q'_j(x) - 
		Q'_i(x) Q_j(x)
		\right) dx\ed 
	\end{equation}
	We can choose $Q_{n}(x)=2^{-n/2}\sqrt{n!}\pi^{1/4}\mathsf{H}_{n}(x)$,
	then $\alpha_{i,j}
	=2^{-(i+j)/2}\sqrt{i!j!}\pi^{1/2}W_{\mathsf{b}}\mathsf{M}_{ij}$ with
	\begin{align}
		&\mathsf{M}_{ij}\equiv \int\mathcal{G}\left(\sqrt{\frac{2}{N}}x\right)e^{-x^2}\left(\mathsf{H}_{i}(x)\mathsf{H}'_{j}(x)-\mathsf{H}'_{i}(x)\mathsf{H}_{j}(x)\right)dx\ed
	\end{align}
	Like what we do for GUE, we find 
	\begin{align}
		&\int\mathcal{G}\left(\sqrt{\frac{2}{N}}x\right)e^{-x^2}\left(\mathsf{H}_{i}(x)\mathsf{H}'_{j}(x)-\mathsf{H}'_{i}(x)\mathsf{H}_{j}(x)\right)dx
		=\langle i|\ii \{\hat{p},\mathcal{G}\left(\sqrt{\frac{2}{N}}\hat{x}\right)\}|j\rangle 
	\end{align}
	where $\ii \left\{ \hat{p},\hat{\mathcal{G}}\right\} =\ii \hat{p}\hat{\mathcal{G}}+\ii \hat{\mathcal{G}}\hat{p},$ and we have made use of $\hat{p}={1\over \ii}{\partial \over \partial x}$. 
	So finally we have 
	\begin{align}
		\langle G\rangle=\sqrt{\frac{\mathsf{\det M}}{2^{N}(2N-1)!!}},~ \mathsf{M}=\mathbb{I}_{2N}\ii \{\hat{p},\hat{\mathcal{G}}\}\mathbb{I}_{2N}\co
	\end{align}
	so that 
	\begin{align}
		{\log \langle G\rangle \over N}={1\over 2N}\left[\log \det \mathsf{M}-N\log2 -\log(2N-1)!!\right]\ed
	\end{align}
	For large $N$ we have 
	\begin{align}
		\frac{\log\langle G\rangle}{N}\approx \frac{1}{2}\left[\frac{\log\det\mathsf{M}}{N}-\left(-1+2\log2+\log N\right)\right]\ed
	\end{align}
	
	In principle, the matrix elements of $\mathsf{M}$ can be computed directly from its definition. 
	However, for the large $N$ limit, we develop a coherent state approach that yields analytical results. 
	Working in the coherent state basis, we find that the operator $\ii \{ \hat{p},\hat{\mathcal{G}}\}$ with $\hat{\mathcal{G}}=e^{w(a+a^\dagger)},w^2\sim {1\over N}$ becomes diagonal
	\begin{equation}
		\begin{aligned}
			\langle \alpha'|\ii \{ \hat{p},\hat{\mathcal{G}}\}|\alpha\rangle
			&\approx \sqrt{2}\left(\alpha-\alpha^{*}\right)e^{w\alpha+w\alpha^*}\delta^2(\alpha-\alpha')\ed
		\end{aligned}
	\end{equation}
	With similar derivations as in the GUE case, we obtain
	\begin{align}
		\frac{\log\det\mathsf{M}}{N} \approx \int_{0}^{\sqrt{2}} u\,du \int_{0}^{2\pi} \frac{d\theta}{\pi} 
		F_{\ii \{\hat{p},\hat{\mathcal{G}}\}}(u,\cos\theta),
	\end{align}
	where the upper bound of integration for $u$ is $\sqrt{2}$ due to the constraint $\text{tr}_R \mathbb{I} = 2N$. The integrand takes the form
	\begin{align}
		F_{\ii \{\hat{p},\hat{\mathcal{G}}\}}(u,\cos\theta) &= \log\left[2\sqrt{2}\ii R u\cos\theta\, \mathcal{G}(\tilde{x})\big|_{\tilde{x}\to 2u\cos\theta}\right] \nn
		&= \log\left[2\sqrt{2}\ii R u\cos\theta\right] + \log\left[\mathcal{G}(\tilde{x})\big|_{\tilde{x}\to 2u\cos\theta}\right],
	\end{align}
	with $R = \sqrt{N}$ as in the GUE case. For the special case $\hat{\mathcal{G}}=1$, this simplifies to
	\begin{align}
		F_{\ii \{\hat{p},\hat{\mathcal{G}}\}}(u,\cos\theta) = \log\left(2^{3/2}R\ii u\cos\theta\right).
	\end{align}
	One can check the normalization is right.  As shown in \cite{Li:2025kpz}, the approach only works for time $t\sim \mathcal{O}(1)$. Thus, we may ask whether the cluster function approach---similar to that used for GUE---can be applied to compute the SFF for GSE in order to study its ramp behavior. Reviewing the treatment in the literature, the key issue lies in calculating \( \mathsf{t}_n \). In the GUE case, we employed the box approximation so that the integration in \( \mathsf{t}_n \) could be performed analytically. However, the GSE case is fundamentally different because the GSE kernel is a matrix. When computing \( \mathsf{t}_n \), a remaining integral persists, where the integrand involves the trace of the product of \( n \) matrices---these being Fourier transforms of the kernel but with different parameters. Moreover, the matrix elements are different functions of these parameters, making analytical analysis rather challenging.
	
	Therefore, we primarily consider numerical methods here. To facilitate rapid computation of the matrix elements, we may simply take \( Q_j(x)=x^j \) to be the simplest polynomial. Thus, we have 
	\begin{align}
		\alpha_{ij}=\left(j-i\right)\int\mathcal{G}(\sqrt{\frac{2}{N}}x)w_{4}(x)x^{i+j-1}dx\ed
	\end{align}
	To calculate SFF of GSE $R$-PSYK$_2$, we define
	\begin{align}
		\mathcal{G}\left(\varepsilon\right)&={1\over D^4}\abs{\sum_{n=0}^{L}d_{n}e^{-itn\left(\varepsilon-\mu\right)}}^4
		=\sum_{k=0}^{2L}\widetilde{g}_k\cos(kt(\varepsilon-\mu))\co
	\end{align}
	then
	\begin{align}
		\alpha_{ij} =\frac{j-i}{2}W_{4}\sum_{\xi=\pm1}\sum_{k=0}^{2L}\widetilde{g}_{k}I_{i+j-1}(\xi kt\sqrt{\frac{2}{N}})e^{-\ii \xi kt\mu}\ed
	\end{align}
	So that one can calculate SFF via
	\begin{align}
		\mathcal{K}(t)=D^{4N}N!\sqrt{\det\boldsymbol{\alpha}}\ed
	\end{align}
	\subsection{GOE}
	Finally, we briefly discuss the case of GOE. Contrary to intuition, the GOE case turns out to be more complicated than GSE. While we can still utilize the expansion method from the literature to reduce the SFF computation to evaluating the determinant of a certain matrix, the key difference lies in the fact that for GOE, the matrix elements (for even $N$) involve a double integral \cite{mehta2004}, in contrast to the single integrals encountered previously:
	\begin{equation}
		\alpha_{ij}= \int_{-\infty}^\infty \int_{-\infty}^\infty \mathcal{G}(\sqrt{\frac{2}{N}}x)\mathcal{G}(\sqrt{\frac{2}{N}}y) w_1(x) w_1(y) R_i(x) R_j(y) g(x - y) dx \, dy,
	\end{equation}
	where
	\begin{equation}
		g(x) =
		\begin{cases} 
			1/2, & x > 0 \\
			-1/2, & x < 0 \\
			0, & x = 0
		\end{cases}\ed
	\end{equation}
	The main difficulty arises from the function $g(x-y)$ in the integrand, which is not an analytic function (polynomial) of $x$ and $y$, making it challenging to obtain an analytic expression for this integral. Consequently, we must resort to numerical integration methods to compute the matrix elements, which significantly increases the computational complexity.
	
	On the other hand, similar to GSE, the kernel for GOE is also a matrix, but with potentially more complicated matrix elements. Therefore, the cluster function approach still fails to yield analytic expressions. In summary, due to the substantial difficulties in both analytic treatment and numerical analysis for GOE, we will not explore this case further in the present work.

	\section{Circular Ensembles}
	\label{sec:CircularE}
	In this section, we investigate the circular ensemble coupling. Three circular ensembles are Hermitian matrices with eigenvalues $e^{\ii \theta_j},j=1,2,\ldots, N$. And the adjoint probability distribution is 
	\begin{align}
		P_{\mathsf{b}}(\theta_1,\cdots,\theta_N)&=\frac{1}{Z_{N,\mathsf{b}}}\prod_{j < k} |e^{i\theta_j}-e^{i\theta_k}|^{\mathsf{b}},\\
		Z_{N,\mathsf{b}}&=(2\pi)^N\frac{\Gamma(\mathsf{b} N/2 + 1)}{(\Gamma(\mathsf{b}/2 + 1))^N}.
	\end{align} 
	where $\mathsf{b}=1,2,4$ corresponding to COE,CUE,CSE. It is found one can obtain an exact formula for the SFF for CUE as shown in \cite{Michael_2024,Ikeda_2025}, where they consider the evolution operator for non-interacting fermions
	\begin{align}
		\mathcal{U}=\exp\left(-i\sum_{i=1}^{N_\mathsf{b}} \theta_i n_i\right)
	\end{align}
	where $n_i=c_i^\dagger c_i$ is the number operator of fermions. 
	And the SFF is defined as 
	\begin{align}
		K_{\mathcal{U}}\left(N,t\right)=\langle|\left(\text{Tr}\mathcal{U}^{t}\right)|^{2}\rangle\ed
	\end{align} 
	The key reason one exact formula can be obtained is due to the simplicity of the cluster function of CUE. Here we choose $\varepsilon_j=\theta_j\in (-\pi,\pi]$ while another choice $\varepsilon_j=e^{\ii \theta_j}$ will lead to a trivial SFF. In this section, we focus on the SFF defined by 
	\begin{equation}
		\begin{aligned}
			\mathcal{K}_{\mathsf{b}}(t)= \left\langle \prod_{j=1}^{N_{\mathsf{b}}} \abs{\sum_{n_{j}=0}^{L} d_{n_j}e^{-i(\varepsilon_{j}-\mu) n_{j}t}}^2 \right\rangle
			=  \int d \theta P_{\mathsf{b}}(\theta) \prod_{j=1}^{N_{\mathsf{b}}} \abs{\sum_{n_{j}=0}^{L} d_{n_j}e^{-i(\varepsilon_{j}-\mu) n_{j}t}}^2
		\end{aligned}
	\end{equation}
	where $N_{\mathsf{b}=4}=2N$ for GSE while $N_{\mathsf{b}=1,2}=N$ for GOE and GSE. Since the effect of chemical potential can be absorbed by a redefinition of $\widetilde{g}_k$. In this section, we only consider the case $\mu=0$.  We can also use the expansion for $\Delta_N$ like Eq.~\eref{eq:Delta-2-exp}. 
	Then one can evaluate SFF by calculating determinant of matrix $\boldsymbol{\alpha}$.
	Similar to the Gaussian ensembles discussed previously, the SFF in principle still admits a determinant expression. However, as demonstrated in \cite{Michael_2024}, we find the cluster function method proves more computationally efficient for this case. The fundamental reason lies in the fact that the kernel of the circular ensembles can be expanded in discrete Fourier series, which allows the integration factors in $\mathsf{r}_n$ to decompose into products of single-variable integrals. Moreover, we observe that the final result can be expressed as the trace of a transfer matrix, thereby establishing a connection with random walks.
	
	Previous theoretical analysis of GUE revealed that the ramp growth rate $C_0$ precisely corresponds to a weighted sum over random walk paths. Therefore, from a technical standpoint, one might expect exact agreement between GUE and CUE results. While it may appear trivial that their SFFs should match in the large $N$ limit since their cluster functions coincide, rigorous verification of this correspondence remains essential.

	\subsection{CUE} 
	We first consider the CUE case, whose kernel is quite simple
	\begin{align}
		\mathsf{K}(\theta_i,\theta_j)=
		\begin{cases}
			\mathsf{R}_1(\theta_i)=\frac{N}{2\pi}&(i = j)\\
			\mathsf{K}(\theta_i - \theta_j)=\frac{\sin\frac{N}{2}(\theta_i - \theta_j)}{2\pi\sin\frac{1}{2}(\theta_i - \theta_j)}&(i\neq j)
		\end{cases}\ed
	\end{align}
	We mainly consider integer time $t\in \mathbb{Z}, t\ge 0$. Using the Fourier representation of kernel
	\begin{align}
		\mathsf{K}(\theta)=\frac{1}{2\pi}e^{-i\frac{N-1}{2}\theta}\sum_{p=0}^{N-1}e^{ip\theta}\co
	\end{align}
	and
	\begin{align}
		\mathsf{R}_{n}(\boldsymbol{\theta}_{n})=\det\left[\mathsf{K}\left(\theta_{j}-\theta_{k}\right)\right]_{j,k=1,\ldots,n}=\sum_{\sigma\in \mathbb{S}_{n}}\text{sgn}\left(\sigma\right)\prod_{j=1}^{n}\mathsf{K}\left(\theta_{j}-\theta_{\sigma(j)}\right)\co
	\end{align} 
	we have 
	\begin{equation}
		\begin{aligned}
			\mathsf{r}_{n}&=\frac{1}{2^{n}}\sum_{\sigma\in\mathbb{S}_{n}}\mathrm{sgn}(\sigma)\sum_{p_{1},\ldots,p_{n}=0}^{N-1}\sum_{\boldsymbol{\xi}_{n}}\int\frac{d\boldsymbol{\theta}_{n}}{(2\pi)^{n}}\mathsf{g_{\boldsymbol{\xi}_{n}}}\exp\left[i\sum_{j=1}^{n}(p_{j}-p_{\sigma(j)}+t\xi_{j})\theta_{j}\right]\\&=\frac{1}{2^{n}}\sum_{\sigma\in\mathbb{S}_{n}}\mathrm{sgn}(\sigma)\sum_{p_{1},\ldots,p_{n}=0}^{N-1}\sum_{\boldsymbol{\xi}_{n}}\mathsf{g_{\boldsymbol{\xi}_{n}}}\prod_{j=1}^{n}\delta(p_{j}-p_{\sigma(j)}+t\xi_{j})\ed
		\end{aligned}
	\end{equation}
	Here $\delta(a)=\delta_{a,0}$ is the Kronecker delta symbol. 
	The permutation sum simplifies to a sum over $\mathbb{S}_n$'s conjugacy classes, which are in bijection with integer partitions of $n$. An integer partition represents $n$ as 
	\begin{equation}
		n = \sum_{i=1}^s \lambda_i m_i
	\end{equation} 
	where $\{\lambda_i\}$ are the parts and $\{m_i\}$ their multiplicities.
	The sign of a permutation is 
	\begin{align}
		\text{sgn}\left(\sigma\right)=(-1)^{\sum_{j=1}^{s}\left(\lambda_{j}-1\right)m_{j}}=(-1)^{n-\sum_j m_j}\ed
	\end{align}
	We denote that a tuple $(\boldsymbol{\lambda}, \boldsymbol{m})$ forms a partition of $n$ by writing $(\boldsymbol{\lambda}, \boldsymbol{m}) \vdash n$, where $\boldsymbol{\lambda} = (\lambda_1, \ldots, \lambda_k)$ are the distinct parts and $\boldsymbol{m} = (m_1, \ldots, m_k)$ their respective multiplicities. To utilize this correspondence, we require the dimension $d(\boldsymbol{\lambda}, \boldsymbol{m})$ of each conjugacy class, given by the well-known formula
	\begin{align}
		d(\boldsymbol{\lambda}, \boldsymbol{m}) = \frac{n!}{\prod_j \left[ m_j! \, \lambda_j^{m_j} \right]}\ed
	\end{align}
	Exactly, we have 
	\begin{align}\label{eq:CUE-rn}
		\mathsf{r}_n &= \sum_{(\boldsymbol{\lambda}, \boldsymbol{m}) \vdash n} (-1)^{n-\sum_j m_j} d(\boldsymbol{\lambda}, \boldsymbol{m}) \prod_j \left[ \Omega_{\lambda_j}(N, t)^{m_j} \right]\nn
		&=(-1)^{n}n!\sum_{(\boldsymbol{\lambda},\boldsymbol{m})\vdash n}\prod_{j}\left(\frac{-\text{Tr}T^{\lambda_{j}}}{2^{\lambda_{j}}\lambda_{j}}\right)^{m_{j}}
	\end{align}
	where we have used $\Omega_{\lambda}(N, t)={1\over 2^{\lambda}}\text{Tr}T^\lambda$ and the transfer matrix 
	\begin{align}
		T_{pp'}\equiv\sum_{\zeta=-L}^L\mathsf{g}_\zeta e^{-\ii t \mu \zeta }\delta\left(p'-p+t\zeta \right)\ed
	\end{align}
	So from Eq.~\eref{eq:G-rn-exp}, we finally have 
	\begin{align}\label{eq:CUE-Kt-T}
		\mathcal{K}(t)=D^{2N}\left[1+\sum_{n=1}^N (-1)^{n}\sum_{(\boldsymbol{\lambda},\boldsymbol{m})\vdash n}\prod_{j}\left(\frac{-\text{Tr}T^{\lambda_{j}}}{2^{\lambda_{j}}\lambda_{j}}\right)^{m_{j}}\right]\ed
	\end{align}
	Using 
	\begin{align}
		\mathsf{T}_{n}(\theta_{1},\dots,\theta_{n})=\sum_{\mathcal{P}(n)}\mathsf{K}(\theta_{1},\theta_{2})\cdots\mathsf{K}(\theta_{n-1},\theta_{n})\mathsf{K}(\theta_{n},\theta_{1}),
	\end{align}
	we have the approximation for large $N$
	\begin{align}
		\mathcal{K}(t)\approx D^{2N} \exp\left[\sum_{n=1}^{N}\frac{(-1)^{n-1}}{n!}\mathsf{t}_{n}\right]= D^{2N}\exp\left[\sum_{n=1}^{N}\frac{(-1)^{n-1}}{n2^{n}}\text{Tr}T^{n}\right]\ed
	\end{align}
	This formula possesses a significant advantage over the exact expression: it allows us to take the logarithm analytically. In contrast, the exact formula Eq.~\eref{eq:CUE-Kt-T} requires computing $\mathsf{r}_n$ terms that largely cancel each other, ultimately yielding extremely small values. Consequently, the exact formula suffers from numerical precision limitations when dealing with large $N$ values. 
	
	However, numerical calculations reveal that this approximation fails for certain values of $m$ in the region where $t/N \ll 1$: precisely the region of our primary interest. Therefore, throughout this work, we employ the exact formula for all calculations. Using \textsf{Mathematica} for numerical computations, one can handle larger $N$ values by increasing the numerical precision of calculations.
	\subsubsection{$z_R(x)=1+mx$}
\begin{figure}[ht]
		\begin{center}
			\includegraphics[width=0.4\textwidth]{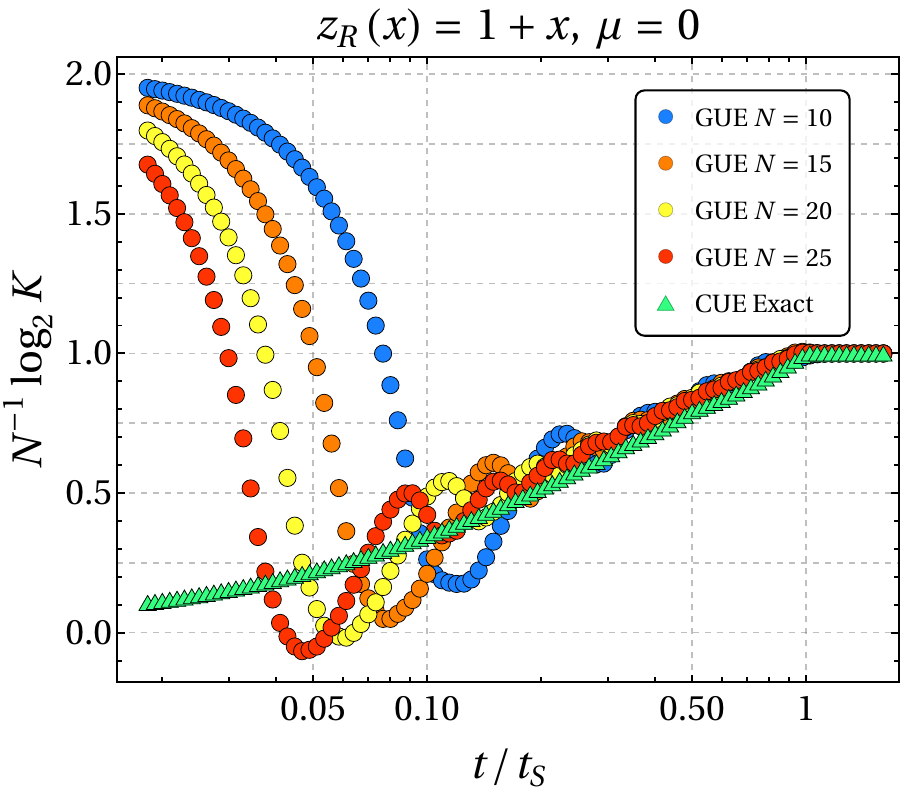}
			\includegraphics[width=0.4\textwidth]{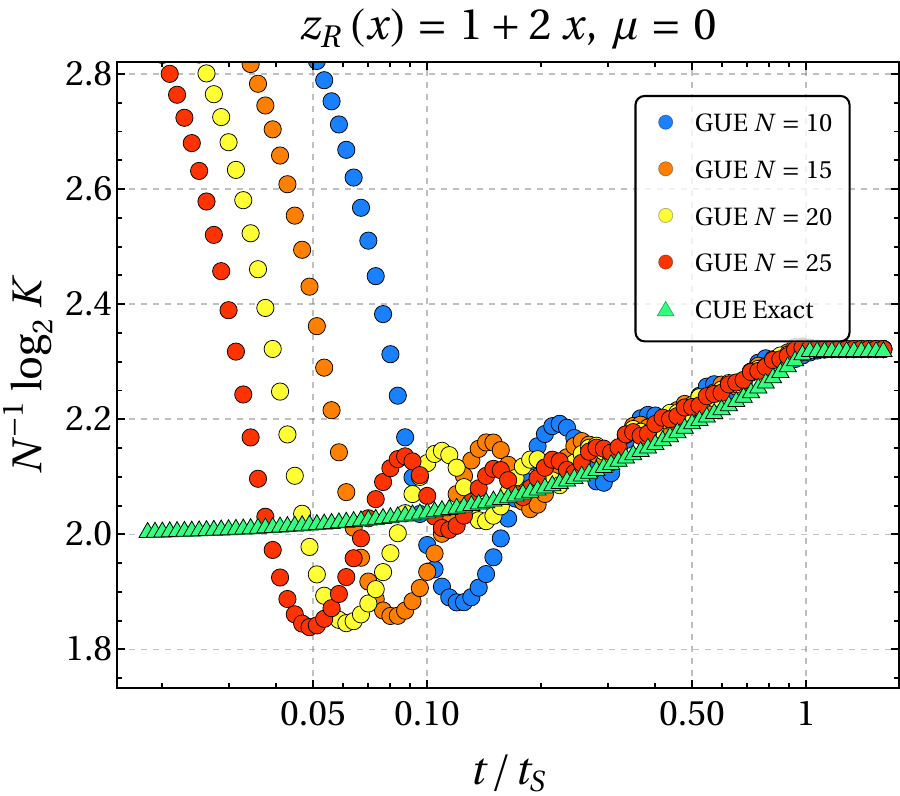}
			\includegraphics[width=0.4\textwidth]{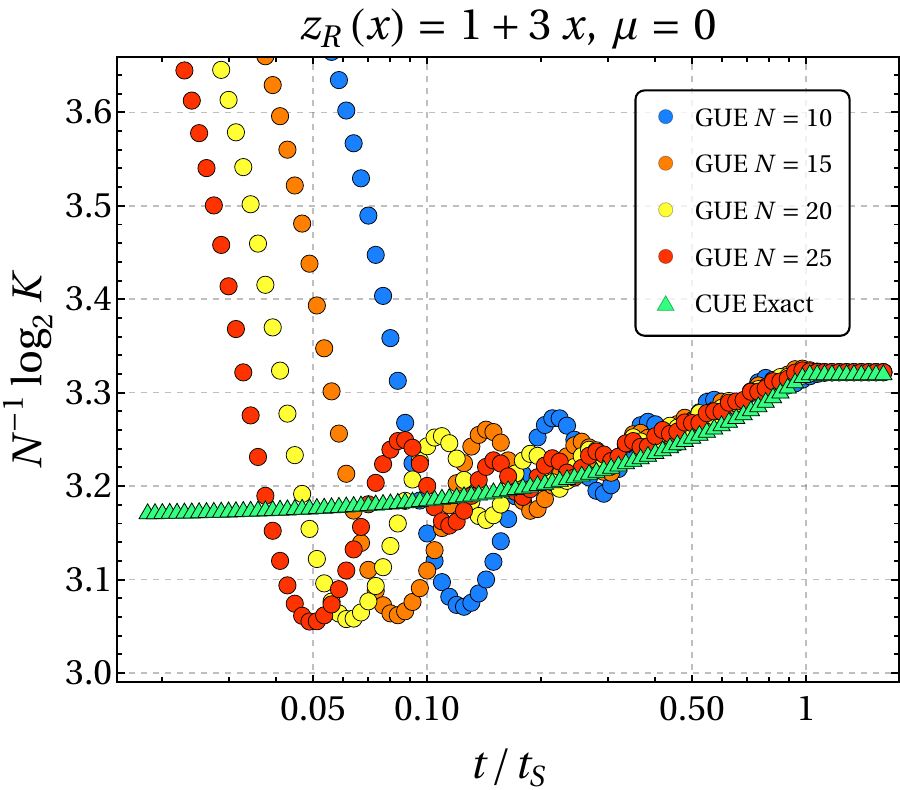}
			\includegraphics[width=0.4\textwidth]{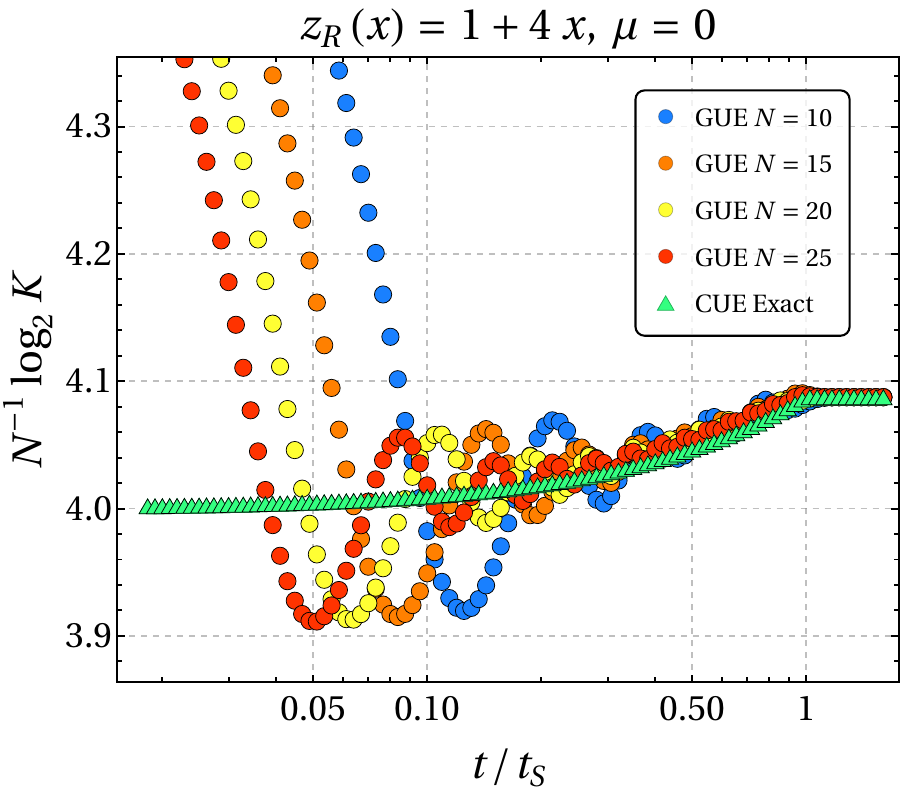}
			\caption{Comparing the SFF of CUE and GUE with $z_R(x) = 1 + mx$.  
			The GUE data is generated by Eq.~\eqref{eq:GUE-SFF-num}, and the CUE data is obtained from Eq.~\eqref{eq:CUE-SFF-recur} with $N = 500$.  
			Scaling time $t_S=2 N$ for GUE and $t_S=N$ for CUE. }
			\label{fig:GUEvsCUE}
		\end{center}
	\end{figure}

\begin{figure}[ht]
		\begin{center}
			\includegraphics[width=0.394\textwidth]{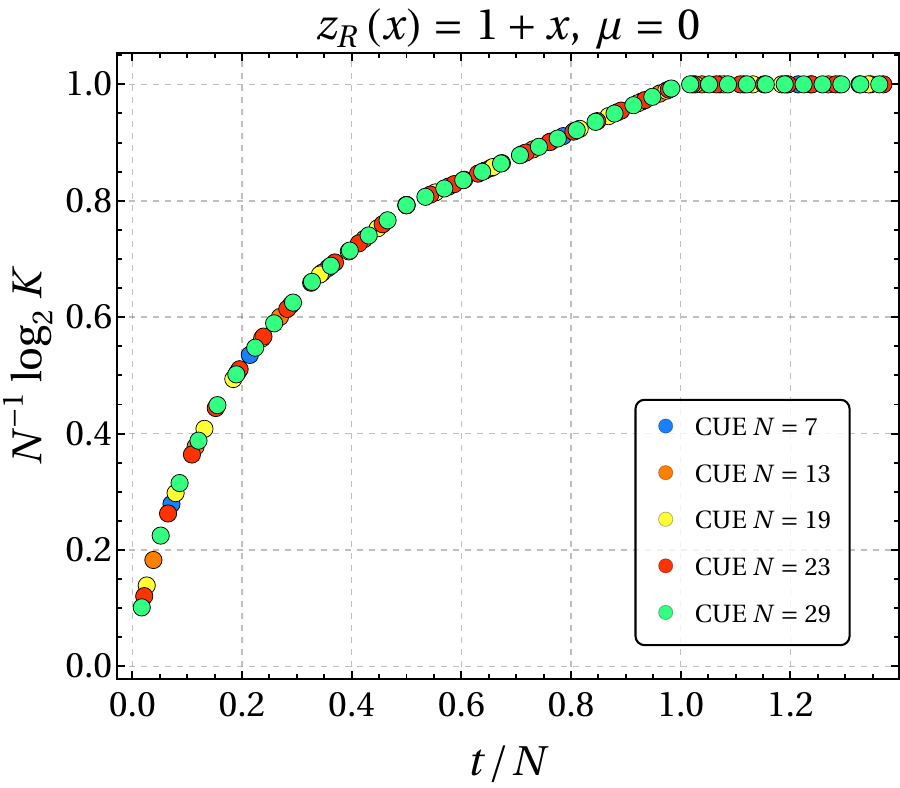}
			\includegraphics[width=0.4\textwidth]{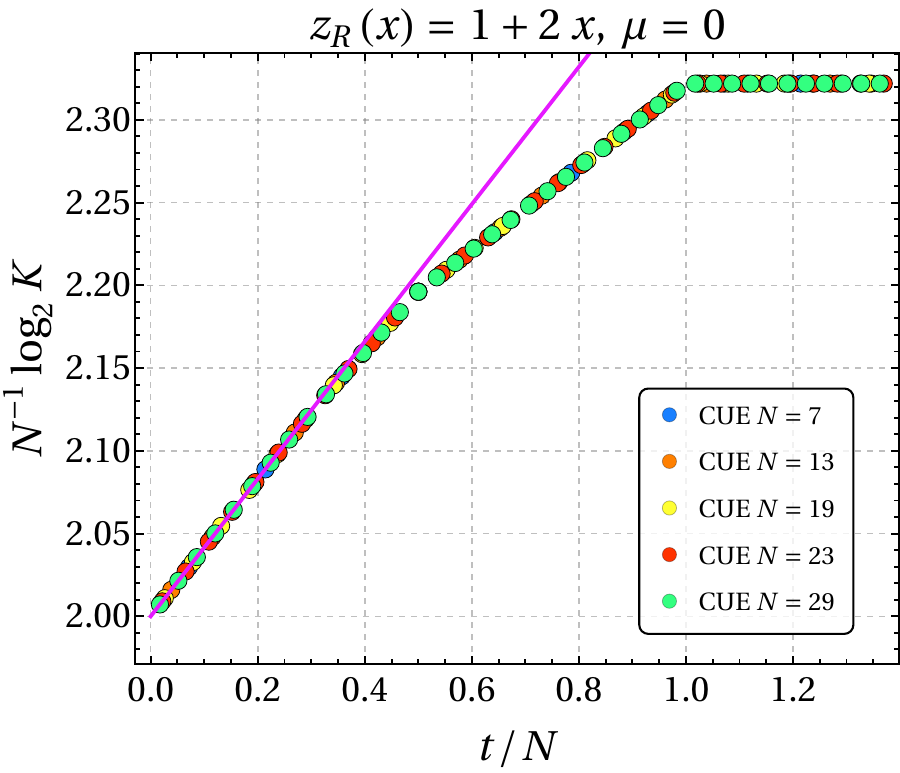}
			\includegraphics[width=0.4\textwidth]{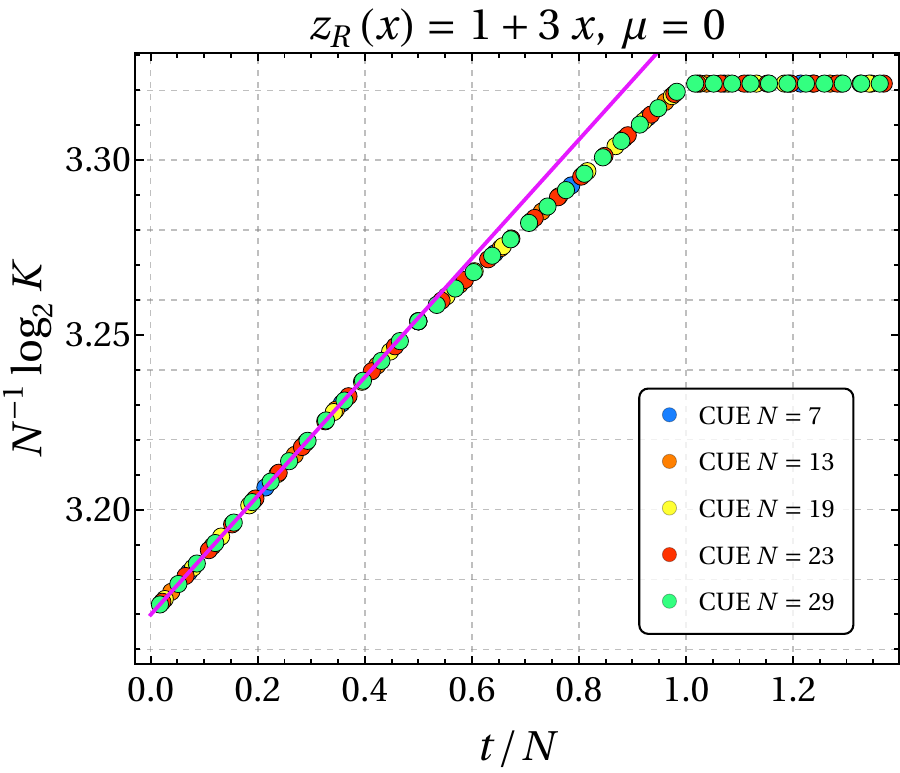}
			\includegraphics[width=0.4\textwidth]{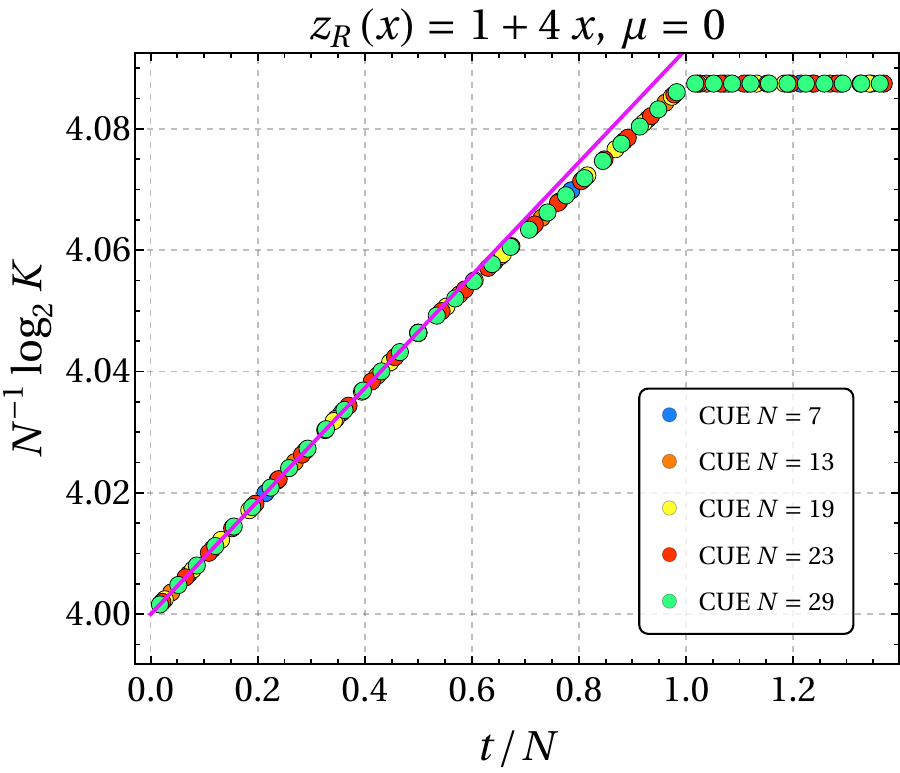}
			\caption{Plot of the SFF for the CUE PSYK model with $z_R(x) = 1 + mx$. The scatter points for the CUE case were generated using the exact formulas in Eq.~\eqref{eq:CUE-SFF-recur} and Eq.~\eqref{eq:CUE-SFF-t1}. The straight lines are plots of Eq.~\eqref{eq:CUE-lines}. }
			\label{fig:CUEvarm}
		\end{center}
	\end{figure}

	For the case $z_R(x)=1+mx,d_0=1,d_1=m$, the SFF can also be analyzed exactly using the cluster function approach. We begin by defining
	\begin{equation}
		\abs{\sum_{n=0}^{1} d_{n}e^{-\ii (\theta -\mu) n t}}^2 = g_{0}\left(1 +  r_1\cos\left(t(\theta-\mu)\right)\right)
	\end{equation}
	where $g_0=1+m^2,r_1=2m/g_0\le 1$. 
	We can evaluate $\mathsf{r}_n$ with the similar method in \cite{Ikeda_2025} 
	\begin{align}
		\mathsf{r}_n = r_1^n\int d{\theta}_n \, \mathsf{R}_n({\theta}_n)  \prod_{i=1}^n \cos(t(\theta_i-\mu))\ed
	\end{align}
	The definition of $\mathsf{r}_n$ is the same as in the reference except for a factor $r_1^n$. So following the derivation in the reference, we finally have
	\begin{equation}
		Q_n(N,t) \equiv r_1^{2n}\sum_{p_1 < \cdots < p_{2n}} \sum_{\sigma \in \mathbb{S}_2^{\otimes n}} \sum_{\boldsymbol{\xi}_{2n}} \prod_{j=1}^{2n} \delta(p_j - p_{\sigma(j)} + t\xi_j)\ed
	\end{equation}
	Here $\xi_j=\pm1$. 
	The combinatorial factor $Q_n(N,t)$ then completely determines the SFF
	\begin{equation}
		\mathcal{K}(N,t) = g_0^N \sum_{n=0}^{\lfloor N/2 \rfloor} \left( \frac{-1}{4} \right)^n Q_n(N,t),
	\end{equation}
	For $N=A t+ r,A=\lfloor N/t \rfloor$ using the Eq. (32) of \cite{Ikeda_2025}
	\begin{align*}
		Q_n(N=At+r,t)=\sum_{n_1,\cdots,n_r=0}^{\lfloor (A+1)/2\rfloor}\sum_{n_{r+1},\cdots,n_t=0}^{\lfloor A/2\rfloor}\left[\prod_{j=1}^r Q_{n_j}(A+1,1)\right]\left[\prod_{k=r+1}^t Q_{n_k}(A,1)\right] \delta\left(n-\sum_{l=1}^tn_l\right),
	\end{align*}
	we still have 
	\begin{align}\label{eq:CUE-SFF-recur}
		\mathcal{K}\left(N,t\right)=\mathcal{K}\left(A+1,1\right)^{r}\mathcal{K}\left(A,1\right)^{t-r}\ed
	\end{align}
	Define $\rho_{\pm}=2-r_{1}^2\pm2\sqrt{1-r_{1}^2}$, one can find the initial SFF is 
	\begin{equation}\label{eq:CUE-SFF-t1}
		\begin{aligned}
			\mathcal{K}(A,1)&=g_{0}^{A}\sum_{n=0}^{\lfloor A/2\rfloor}\left(\frac{-1}{4}\right)^{n}\binom{A-n}{n}r_{1}^{2n}\\&=\begin{cases}
				A+1, & r_{1}=1\\
				\frac{g_{0}^{A}}{2^{A+1}}\left(\rho_{+}^{A/2}+\rho_{-}^{A/2}+\frac{\rho_{+}^{A/2}-\rho_{-}^{A/2}}{\sqrt{1-r_{1}^{2}}}\right), & r_{1}<1,A\ \text{even}\\
				\frac{g_{0}^{A}}{2^{A}}\left[\rho_{-}^{\frac{A-1}{2}}+\rho_{+}^{\frac{A-1}{2}}+\frac{\left(2-r_{1}^{2}\right)}{2\sqrt{1-r_{1}^{2}}}\left(\rho_{+}^{\frac{A-1}{2}}-\rho_{-}^{\frac{A-1}{2}}\right)\right], & r_{1}<1,A\ \text{odd}
			\end{cases}\ed
		\end{aligned}
	\end{equation}
	\begin{figure}[ht]
		\begin{center}
			\includegraphics[width=0.387\textwidth]{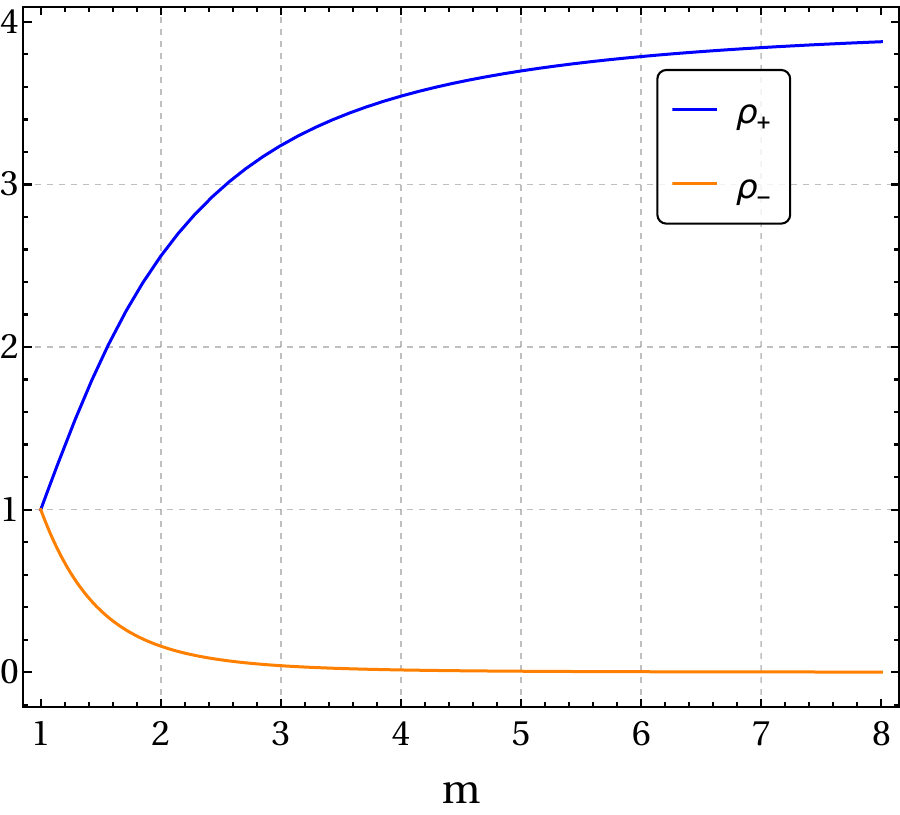}
			\includegraphics[width=0.4\textwidth]{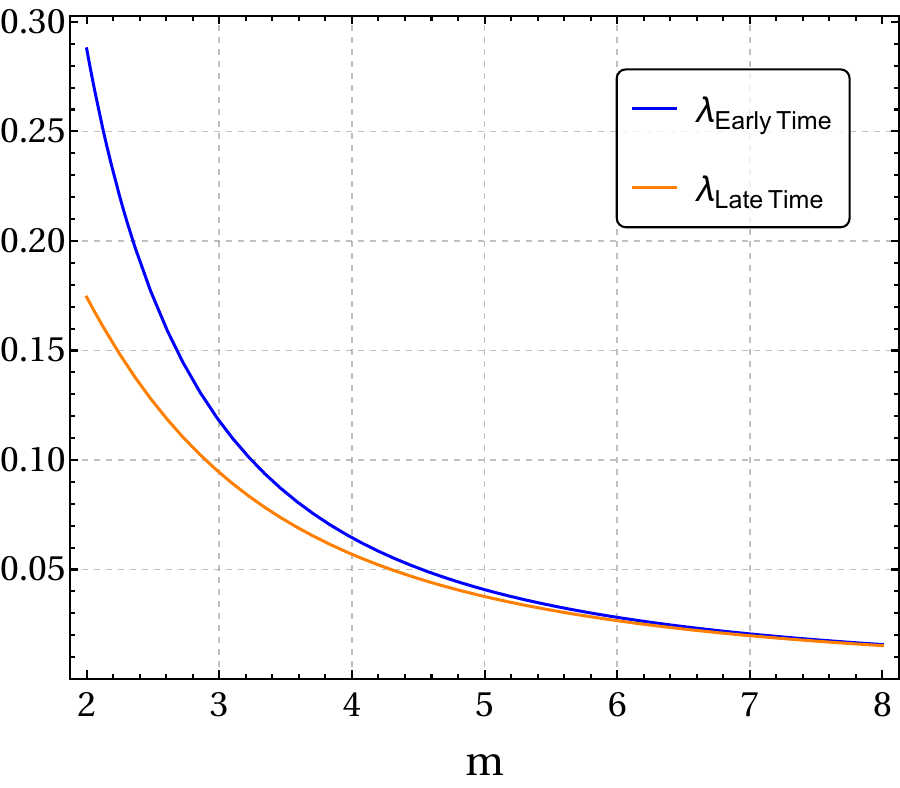}
			\caption{Plots of $\rho_{\pm}$ and the growth rates for Example A. }
			\label{fig:rho-lam}
		\end{center}
	\end{figure}
	As depicted in Fig.~\ref{fig:rho-lam}, we have $0<\rho_- \le 1\le \rho_+\le 4$.  So for large $A$,  we simply have 
	\begin{align}
		\mathcal{K}\left(A,1\right)\stackrel{A\gg1}{=}\begin{cases}
			A+1, & r_{1}=1\\
			\frac{g_{0}^{A}\rho_{+}^{A/2}}{2^{A+1}}\left(1+\frac{1}{\sqrt{1-r_{1}^{2}}}\right), & r_{1}<1,A\ \text{even}\\
			\frac{g_{0}^{A}\rho_{+}^{(A-1)/2}}{2^{A}}\left(1+\frac{2-r_{1}^{2}}{2\sqrt{1-r_{1}^{2}}}\right), & r_{1}<1,A\ \text{odd}
		\end{cases}\ed
	\end{align}
	For $t> N$, we have $A=0, r=N$, so that we have the plateau behavior 
	\begin{align}
		\mathcal{K}(N,t> N)=\mathcal{K}\left(1,1\right)^{N}=\begin{cases}
			2^{N} & r_{1}=1\\
			g_0^N& r_{1}<1
		\end{cases}\ed
	\end{align}
	Before reaching the plateau, the SFF consists of a sequence of exponential ramps with growth rates that depend on \(\lfloor N/t \rfloor \). To see this, fix \( L \) and make a list of its divisors 
	\begin{align}
		1 = t_0 < t_1 < \cdots < t_{N-1} = N
	\end{align}
	where $(N-k)t_k=N$. For any time $t_{j-1} < t \leq t_{j}$, we have $A= A_j\equiv N-j$ is fixed. Then 
	\begin{align}
		\mathcal{K}_N(t_{j-1} < t \leq t_{j})=\mathcal{K}\left(A_{j}+1,1\right)^{t_{j-1}-A_{j}\delta t}\mathcal{K}\left(A_{j},1\right)^{\left(A_{j}+1\right)\delta t}\equiv \mathcal{K}(t_{j-1})e^{\lambda_j \delta t}\co
	\end{align}
	so that the growth rate is 
	\begin{align}
		\lambda_j=\left(A_{j}+1\right)\log\mathcal{K}\left(A_{j},1\right)-A_{j}\log\mathcal{K}\left(A_{j}+1,1\right)\ed
	\end{align}
	As shown in \cite{Liao_2020}, the fermionic $\text{SYK}_2$ model with GUE couplings exhibits an exponential ramp in the time regime $1 \ll t \ll N$, with growth rate $\lambda$ given by
	\begin{align}
		\lambda_{\text{GUE}}=\frac{1}{4}\left(\ln\frac{N}{8}+\gamma_{E}\right) \sim \mathcal{O}(\log N)\ed
	\end{align} 
	Here we can analytically check it by taking the limit $A=\lfloor N /t \rfloor \gg 1$. For $r_1=1$, we have 
	\begin{align}
		\lambda_{\text{CUE}}=\left(A+1\right)\log\left(A+1\right)-A\log\left(A+2\right)= \log A -1 +\mathcal{O}(1/A)\sim \mathcal{O}(\log N)\co
	\end{align}
	which is consistent with the GUE case. While for $r_1<0$, it is shown that $\lambda_{\text{GUE}}^{\infty}=\frac{1}{2}\log\left(\frac{m^{2}}{m^{2}-1}\right)$ for large $N$ limit. Here we consider large $A$ limit, and we find 
	\begin{align}
		\lambda_{\text{CUE}}^{\infty} = \lim_{A\to \infty} \left(A+1\right)\log\mathcal{K}\left(A ,1\right)-A \log\mathcal{K}\left(A+1,1\right)= \log\left(\frac{m^{2}}{m^{2}-1}\right)\ed
	\end{align}
	They are not the same, but we exactly have 
	\begin{align}\label{eq:lambda-eq}
		2\lambda_{\text{GUE}}^{\infty}=\lambda_{\text{CUE}}^{\infty}\ed
	\end{align}
	For $t/N\ll 1$, we have an approximation for $\mathcal{K}(t)$ for $m>1$
	\begin{align}\label{eq:CUE-lines}
		\frac{1}{N}\log_{2}\mathcal{K}(N,t)=\log_{2}\left(\frac{m^{2}}{m^{2}-1}\right)\frac{t}{N}+\log_2\left({1\over 2}g_0\sqrt{\rho_+}\right)\ed
	\end{align}
	Moreover, for any time $1/2<t/N<1$, we have  
	\begin{align}
		\lambda=2\log\mathcal{K}\left(1,1\right)-\log\mathcal{K}\left(2,1\right)=\log\left(\frac{\left(m^{2}+1\right)^{2}}{\left(m^{2}+1\right)^{2}-m^{2}}\right)\sim \mathcal{O}(1)\ed
	\end{align}
	Notice that the SYK$_2$ with GUE coupling reach the plateau at time $t_*=2N$ while the CUE case gives $t_*=N$. One can guess their SFFs coincide up to a time scaling. To be consistent with Eq. \eref{eq:lambda-eq} and the plateau behavior, we propose
	\begin{align}\label{eq:scale-guess}
		\mathcal{K}_{\text{GUE}}(2t)\approx\mathcal{K}_{\text{CUE}}(t)\ed
	\end{align} 
	In \cite{Michael_2024}, the authors scaled time by relevant single-particle Heisenberg time, namely
	\begin{align}
		\mathcal{K}_{\text{GUE}}\left({\pi\over \sqrt{2}}t\right)\approx \mathcal{K}_{\text{CUE}}(t)\ed
	\end{align}
	Notice that ${\pi\over \sqrt{2}}\approx 2.22144$ is near to $2$. Here we adopt the scaling guess Eq.~\eref{eq:scale-guess} for it matches both the growth rate and the plateau place exactly.  Use this approximation, we can determine the exponential ramp region for fermionic SYK$_2$ by imposing Eq.~\eref{eq:lambda-eq} for $r_1=1$, so that we have 
	\begin{align}
		\frac{1}{2}\left(\ln\frac{N}{8}+\gamma_{E}\right)=\log A -1\ed 
	\end{align} 
	Then we have $A=e^{1+\gamma_{E}/2-\log2}\sqrt{N}$ with typical time $t=N/A\sim \sqrt{N}$. The perfect agreement between GUE and CUE is demonstrated in Figs.~\ref{fig:GUEvsCUE} and~\ref{fig:CUEvarm}. Furthermore, we observe that the growth rate exhibits similar trends at both early times $t/N\ll 1$ and late times $1/2<t/N<1$, as illustrated in  Fig.~\ref{fig:rho-lam}. 
	\subsubsection{$z_R(x)=1+mx+x^2$}
\begin{figure}[ht]
		\begin{center}
			\includegraphics[width=0.48\textwidth]{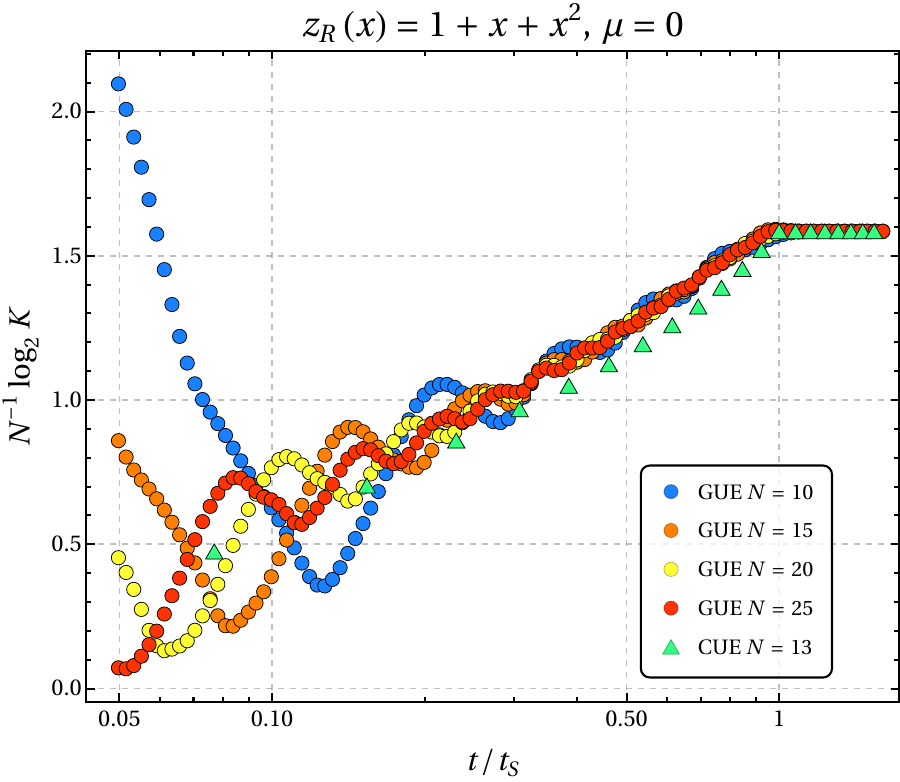}
			\includegraphics[width=0.48\textwidth]{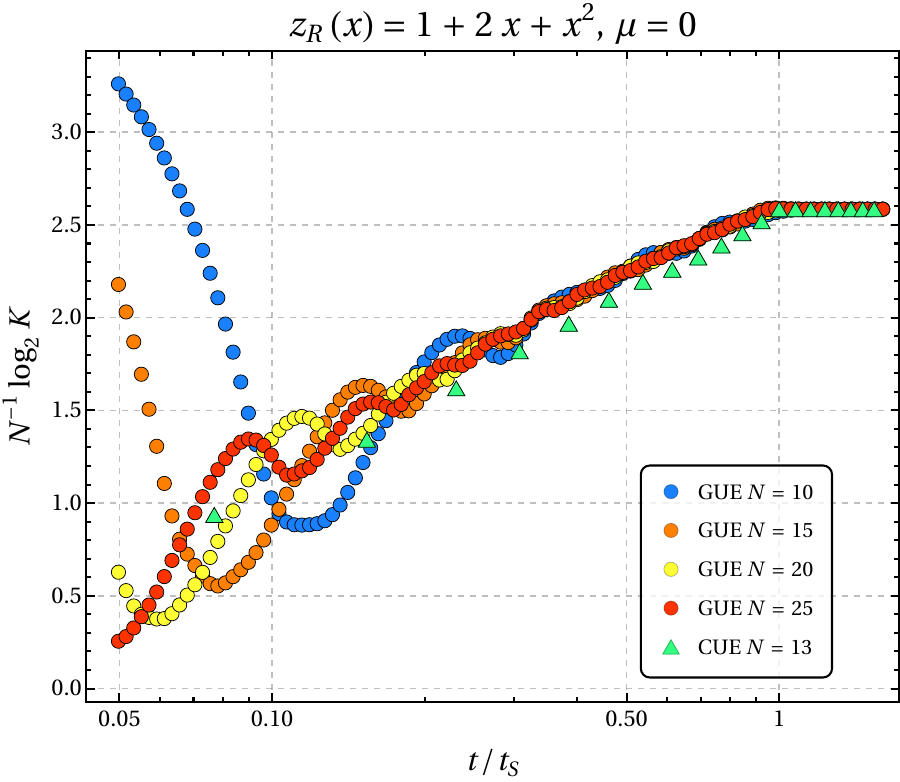}
			\includegraphics[width=0.48\textwidth]{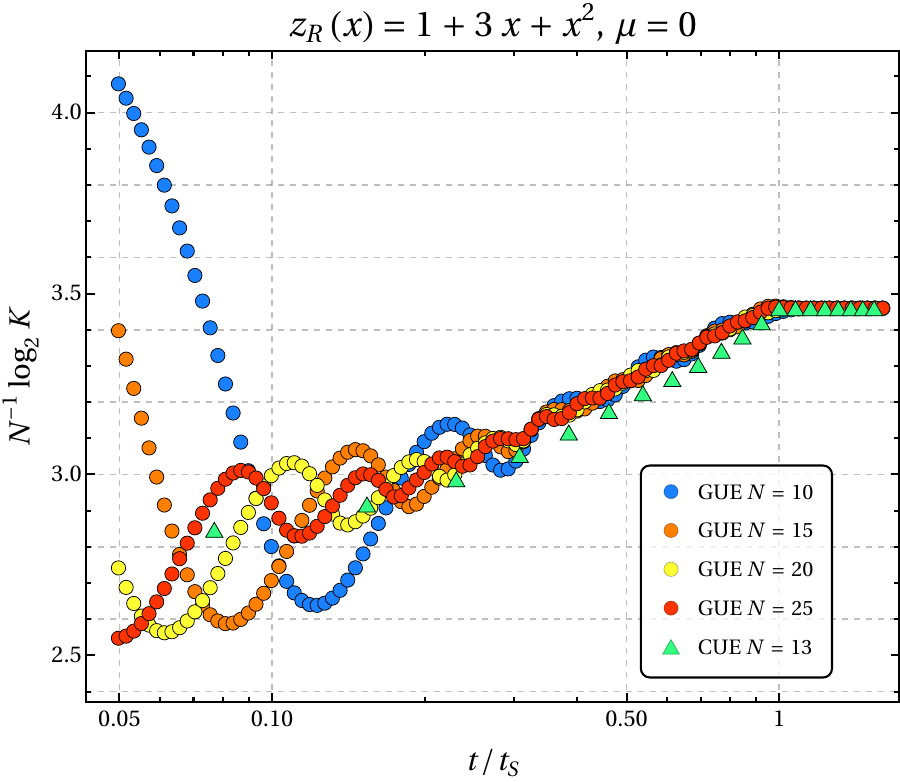}
			\includegraphics[width=0.48\textwidth]{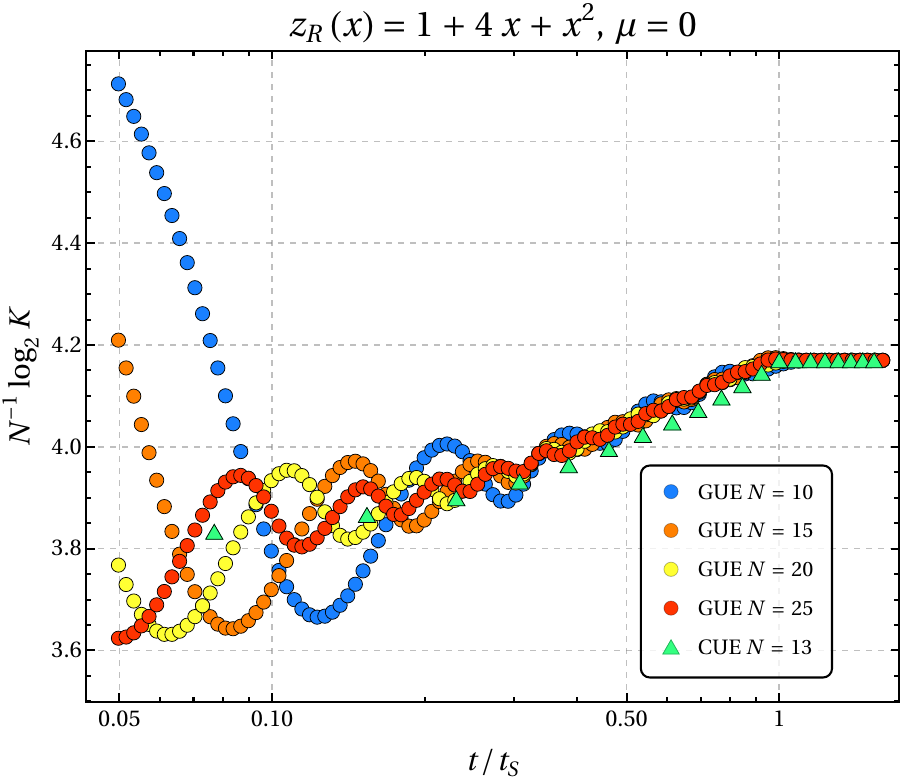}
			\caption{Comparing SFFs of CUE and GUE with $z_R(x)=1+mx+x^2$. Here data of GUE is generated by Eq.~\eref{eq:GUE-SFF-num} and the data of CUE is calculated numerically via Eq.~\eref{eq:CUE-rn}. Scaling time $t_S=2 N$ for GUE and $t_S=N$ for CUE. }
			\label{fig:GUEvsCUE_B}
		\end{center}
	\end{figure}

	For Example B, we do not obtain an analytical representation of its SFF. As shown in \cite{Li:2025kpz}, we know that its SFF have an exponential ramp in the time regime $(N/C_0)^{2/5}\ll t \ll N$ with growth rate $C_0$. The numerical result for $C_0$ in the large $N$ limit is represented in Table~\ref{table:C0-GUE}. 
	But a numerical check is still lacked. Here we checked it by CUE. We want to prove $C_0$ is nothing but half of the growth rate of CUE SFF at time $t/N\ll 1$ in the large $N$ limit, as Eq.~\eref{eq:lambda-eq} shows for the case $z_R(x)=1+mx$. We first need to show that CUE is a good approximation for GUE. As depicted in Fig.~\ref{fig:GUEvsCUE_B}, although the two SFFs are not coincides with each other, but they share the same trend.  
	\begin{figure}[ht]
		\begin{center}
			\includegraphics[width=0.48\textwidth]{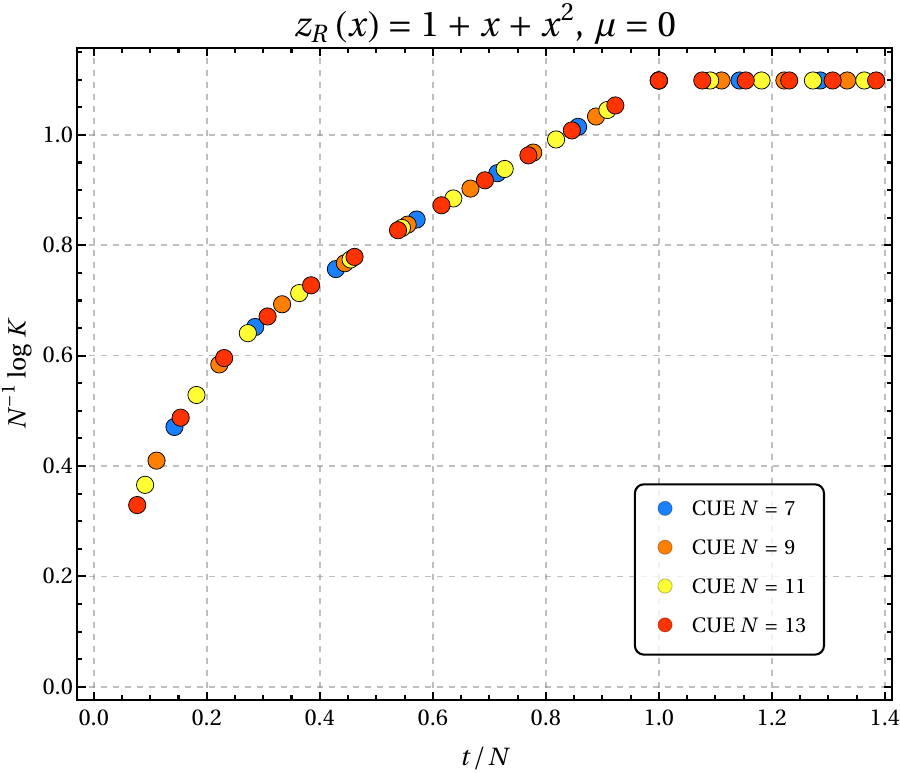}
			\includegraphics[width=0.48\textwidth]{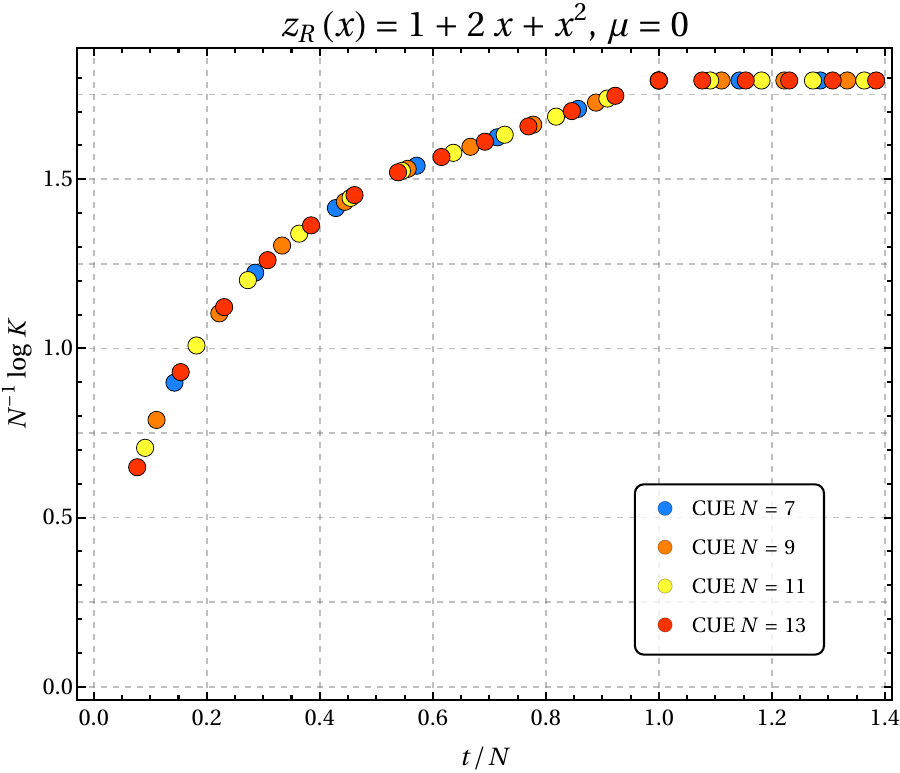}
			\includegraphics[width=0.48\textwidth]{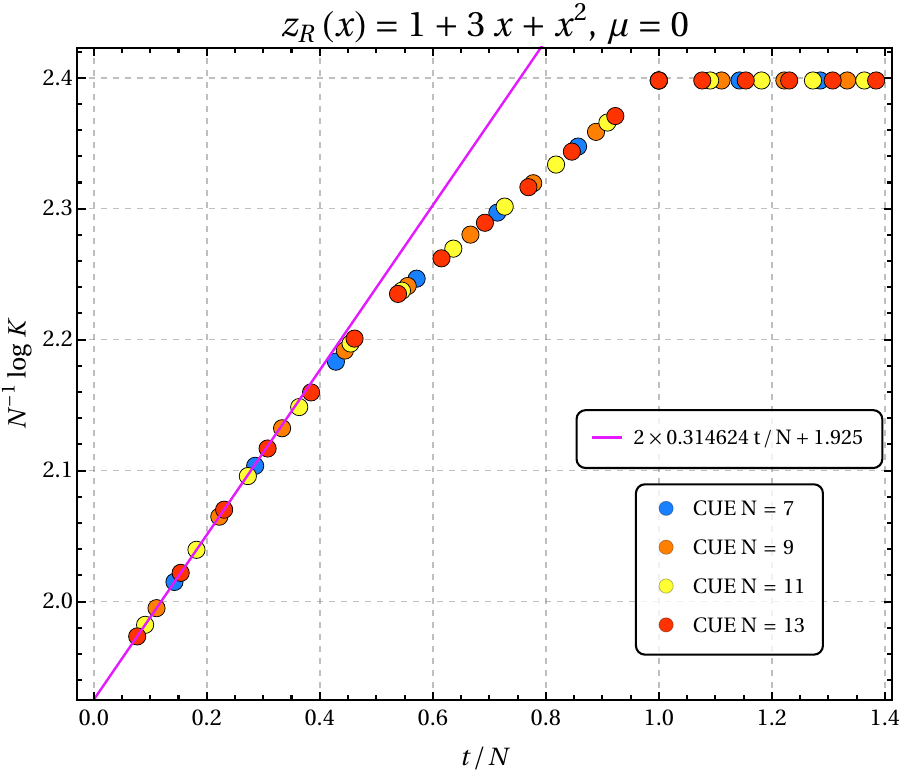}
			\includegraphics[width=0.48\textwidth]{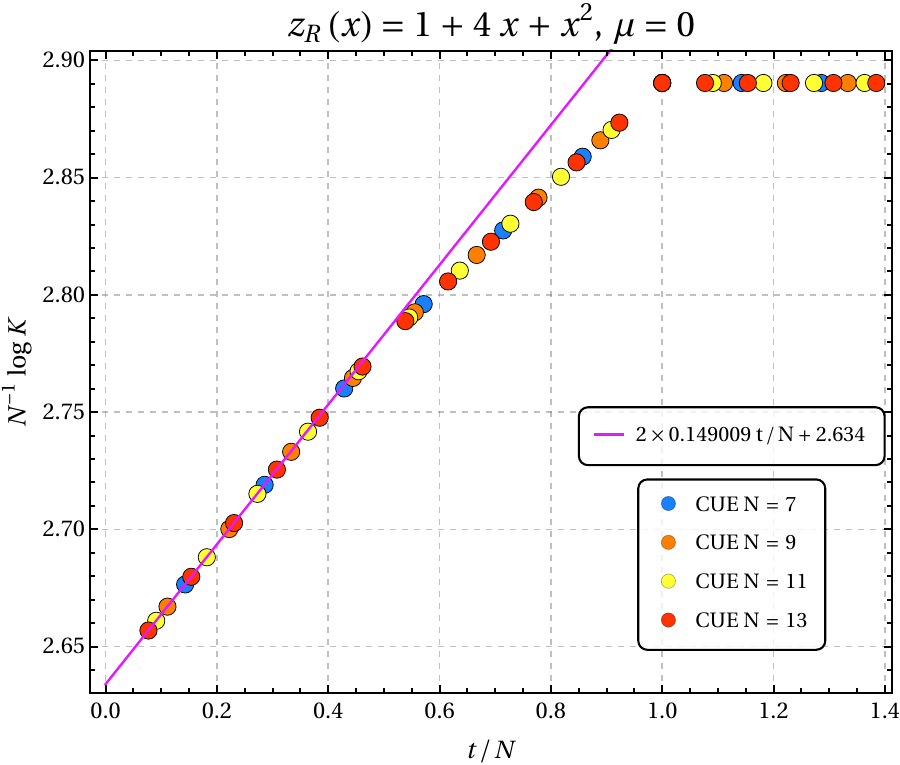}
			\caption{SFF for CUE coupling with $z_R(x)=1+mx+x^2$. 
				The CUE data (points) is generated using Eq.~\eqref{eq:CUE-Kt-T}, 
				with fitted straight lines having slopes $2C_0$ as specified 
				in Table~\ref{table:coeff}. The theoretical GUE results (solid lines) 
				show excellent agreement with CUE numerical data in the region $t/N\ll 1$. 
				For $m=1,2$ cases, the expected infinite slope behavior is consistent 
				with the trend observed in the CUE numerical results.  }
			\label{fig:CUEexBvarm}
		\end{center}
	\end{figure}
	Then we move to discuss the growth rate.

	Since the ramp appears in the region $t \ll N$ for the GUE case, we verify the correctness of our results by examining the plot of $\frac{1}{N} \ln \mathcal{K}(t)$ versus $t/N$. At $t/N = 0$, we expect the curve to have a slope of $2C_0$. As shown in Fig.~\ref{fig:CUEexBvarm}, the slope diverges for $m = 1, 2$ at $t/N = 0$, while the cases $m = 3, 4$ agree well with the GUE results.
	
	We can also consider the non-Hermitian case where $\varepsilon = e^{i\theta}$. Although SFF can be formally defined in this scenario, one finds that it becomes time-independent. Notably, 
	\begin{equation}
		\begin{aligned}
			&\int d\theta_{1}\ldots d\theta_{n}\,\mathsf{K}(\theta_{1},\theta_{2})\ldots\mathsf{K}(\theta_{n},\theta_{1})e^{it\sum_{i=1}^{n}e^{i\theta_{i}}\zeta_{i}}\\=&\frac{1}{\left(2\pi\right)^{n}}\int d\theta_{1}\ldots d\theta_{n}\,\sum_{p_{1}=0}^{N-1}e^{ip_{1}\left(\theta_{1}-\theta_{2}\right)}\sum_{p_{2}=0}^{N-1}e^{ip_{2}\left(\theta_{2}-\theta_{3}\right)}\ldots\sum_{p_{n}=0}^{N-1}e^{ip_{n}\left(\theta_{n}-\theta_{1}\right)}e^{it\sum_{i=1}^{n}e^{i\theta_{i}}\zeta_{i}}\\=&\frac{1}{\left(2\pi i\right)^{n}}\oint_{|z_i|=1}\frac{dz_{1}}{z_{1}}\frac{dz_{2}}{z_{2}}\ldots\frac{dz_{n}}{z_{n}}\,\sum_{\left\{ p_{1},p_{2},\ldots,p_{n}\right\} =0}^{N-1}\left(\frac{z_{1}}{z_{2}}\right)^{p_{1}}\left(\frac{z_{2}}{z_{3}}\right)^{p_{1}}\cdots\left(\frac{z_{n}}{z_{1}}\right)^{p_{n}}e^{it\sum_{j=1}^{n}\zeta_{j}z_{j}}\\=&N\ed
		\end{aligned}
	\end{equation}
	Using $\sum_{k=-L}^L \mathsf{g}_k=0$, we have $\mathsf{t}_n=0$, $\mathcal{K}(t)=D^{2N}$. 
	\subsection{CSE}
	Notice the spectrum of CSE is double-fold, thus 
	\begin{equation}
		\begin{aligned}
			\mathcal{K}_{\mathsf{b}}(t)= \left\langle \prod_{j=1}^{N} \abs{\sum_{n_{j}=0}^{L} d_{n_j}e^{-i(\varepsilon_{j}-\mu) n_{j}t}}^4 \right\rangle
			=  \int d \theta P_{\mathsf{b}}(\theta) \prod_{j=1}^{N} \abs{\sum_{n_{j}=0}^{L} d_{n_j}e^{-i(\varepsilon_{j}-\mu) n_{j}t}}^4
		\end{aligned}\ed
	\end{equation}
	Its kernel has a matrix form
	\begin{align}
		\mathsf{K}_{N,4}(\theta) = \frac{1}{2} \begin{pmatrix}
			S_{2N}(\theta) & D_{2N}(\theta) \\
			I_{2N}(\theta) & S_{2N}(\theta)
		\end{pmatrix}
	\end{align}
	where 
	\begin{align}\label{eq:SDI-def}
		S_{2N}(\theta) \equiv \frac{1}{2\pi} \frac{\sin(N\theta)}{\sin(\theta/2)},~
		D_{2N}(\theta) \equiv \frac{d}{d\theta} S_{2N}(\theta), 
		I_{2N}(\theta) \equiv \int_0^\theta d\phi \, S_{2N}(\phi).
	\end{align}
	Like the CUE case, it is convenient to represent each element of the kernel as a Fourier series
	\begin{equation}
		S_{2N}(\theta) = \frac{1}{2\pi} \sum_p e^{ip\theta},~D_{2N}(\theta) = \frac{i}{2\pi} \sum_p p \, e^{ip\theta},~I_{2N}(\theta) = \frac{1}{2\pi i} \sum_p \frac{e^{ip\theta}}{p}.
	\end{equation}
	Here and throughout this subsection, the sums over \( p \) are taken over the half-integers,
	\begin{equation}
		p \in \left\{ -\frac{2N - 1}{2}, -\frac{2N - 1}{2} + 1, \ldots, \frac{2N - 1}{2} \right\}.
		\label{eq:pvalues}
	\end{equation}
	Then one find a compact notation for the kernel
	\begin{equation}
		K^{\alpha,\beta}_{N,4}(\theta) = \frac{i^{\beta-\alpha}}{4\pi} \sum_p p^{\beta-\alpha} e^{ip\theta}
		\label{eq:Kernel}
	\end{equation}
	where \(\alpha, \beta = 0, 1\) specify matrix elements of the kernel. For any permutation consisting of cycles, we have $\sigma=\sigma_1\sigma_2\ldots \sigma_{n_c}$, and we denote the length of each cycle as $|\sigma_j|$ with elements $a_{j,1}\to a_{j,2}\to a_{j,3},\ldots a_{j,|n_j|}$, we have
	\begin{align}
		\mathsf{R}_n(\boldsymbol{\theta}_n) &= \sum_{\sigma\in\mathbb{S}_n}(-1)^{n-n_c(\sigma)} 
		\prod_{j=1}^{n_c(\sigma)}\left\{\frac{1}{2}\mathrm{Tr}\left[\prod_{k=1}^{|\sigma_j|}\mathsf{K}_{N,4}(\theta_{a(j,k)}-\theta_{a(j,k+1)})\right]\right\},
	\end{align}
	so that we can calculate 
	\begin{align}
		\mathsf{r}_{n}&= \int d\varepsilon_{1}\ldots d\varepsilon_{n}\,\mathsf{R}_{n}(\varepsilon_{1},\dots,\varepsilon_{n})\prod_{i=1}^{n}F(\varepsilon_{i},t)\ed
	\end{align}
	$F(\varepsilon,t)$ in above equation should be defined as
	\begin{align}
		\abs{\sum_{n=0}^L d_n e^{in t (\varepsilon-\mu)}}^4=D^4(1+F(\varepsilon,t)), F(\varepsilon,t)\equiv {1\over 2}\sum_{k=-2L}^{2L} \mathsf{g}_k e^{ik(\varepsilon-\mu)}\ed
	\end{align} 
	The expressions of $\mathsf{g}_k$ are different with CUE case. Similar to the derivation of CUE, the exact formula of CSE SFF is 
	\begin{align}\label{eq:CSE-Kt-T}
		\mathcal{K}(t)=D^{4N}\left[1+\sum_{n=1}^N (-1)^{n}\sum_{(\boldsymbol{\lambda},\boldsymbol{m})\vdash n}\prod_{j}\left(\frac{-\text{Tr}\mathcal{T}^{\lambda_{j}}}{2^{2\lambda_{j}+1}\lambda_{j}}\right)^{m_{j}}\right]
	\end{align}
	where the $4N\times 4N$ transfer matrix is 
	\begin{align}
		\mathcal{T}_{p\alpha,p'\alpha'}=p^{\alpha-\alpha'}\sum_{\zeta}\mathsf{g}_{\zeta}\delta(p-p'+t\zeta_{j})\ed
	\end{align}
	For large $N$, we can calculate SFF via $\mathsf{t}_n$. Using the $n$-point cluster function 
	\begin{align}
		\mathsf{T}_n(\varepsilon_1, \dots, \varepsilon_n) = \sum_{\mathcal{P}(n)} {1\over 2}\text{Tr}\left[\mathsf{K}(\varepsilon_1, \varepsilon_2) \cdots \mathsf{K}(\varepsilon_{n-1}, \varepsilon_n) \mathsf{K}(\varepsilon_n, \varepsilon_1)\right],
	\end{align}
	we finally have the approximation 
	\begin{align}
		\mathcal{K}(t)\approx \exp\left[\sum_{n=1}^{N}\frac{(-1)^{n-1}}{n!}\mathsf{t}_{n}\right]=\exp\left[\sum_{n=1}^{N}\frac{(-1)^{n-1}}{n2^{2n+1}}\text{Tr}\mathcal{T}^{n}\right]\ed
	\end{align}
	\begin{figure}[ht]
		\begin{center}
			\includegraphics[width=0.39\textwidth]{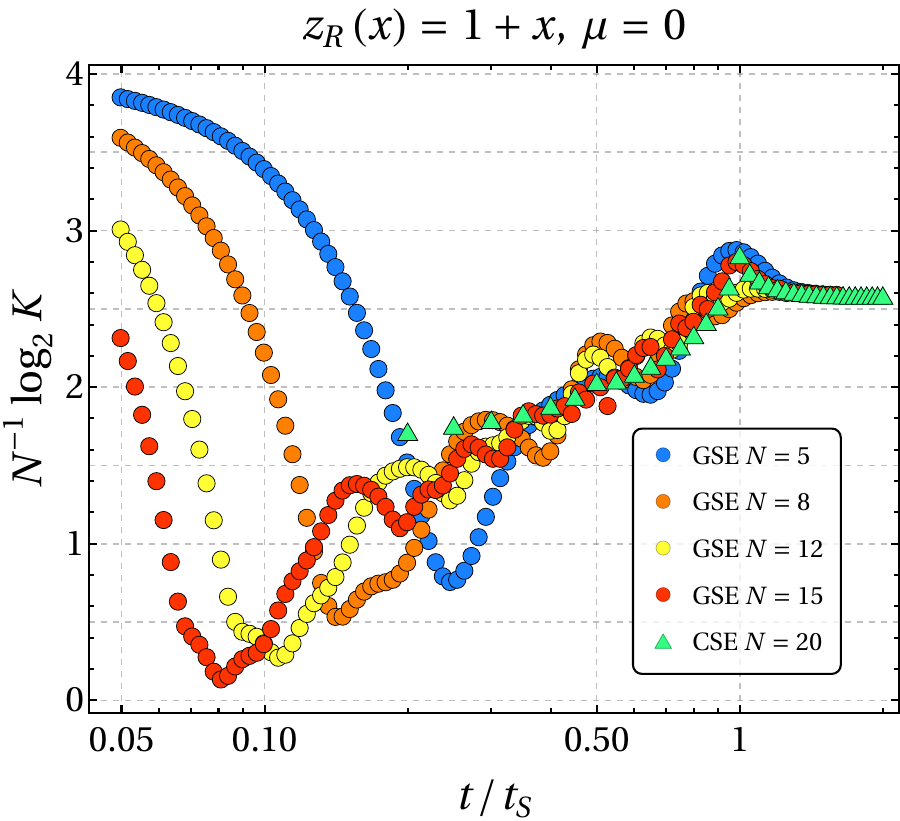}
			\includegraphics[width=0.4\textwidth]{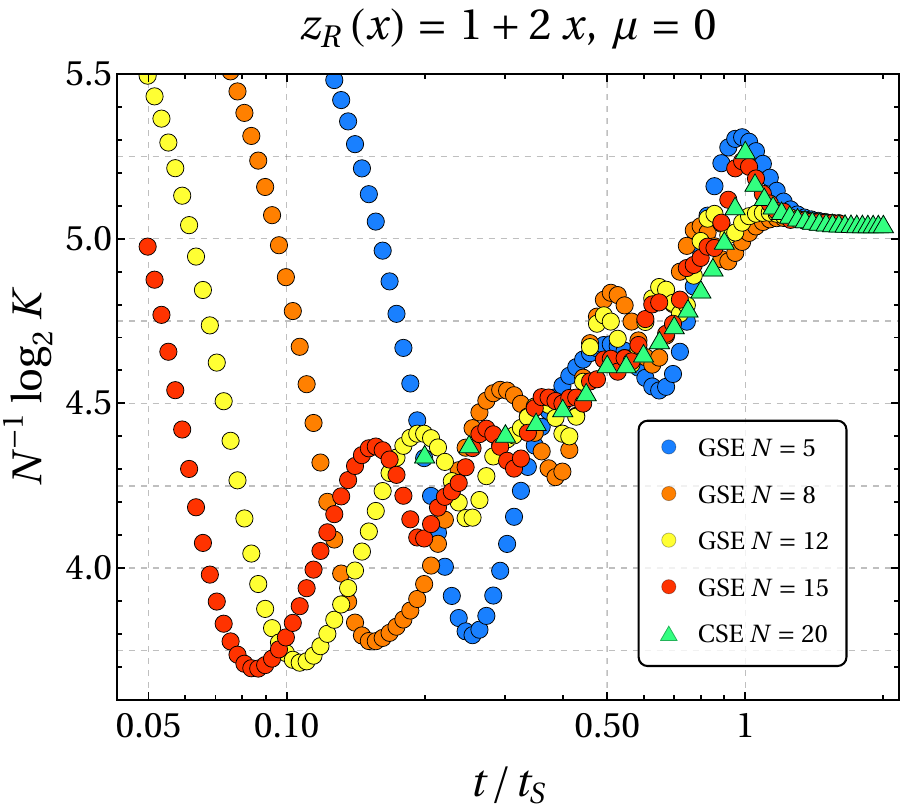}
			\includegraphics[width=0.4\textwidth]{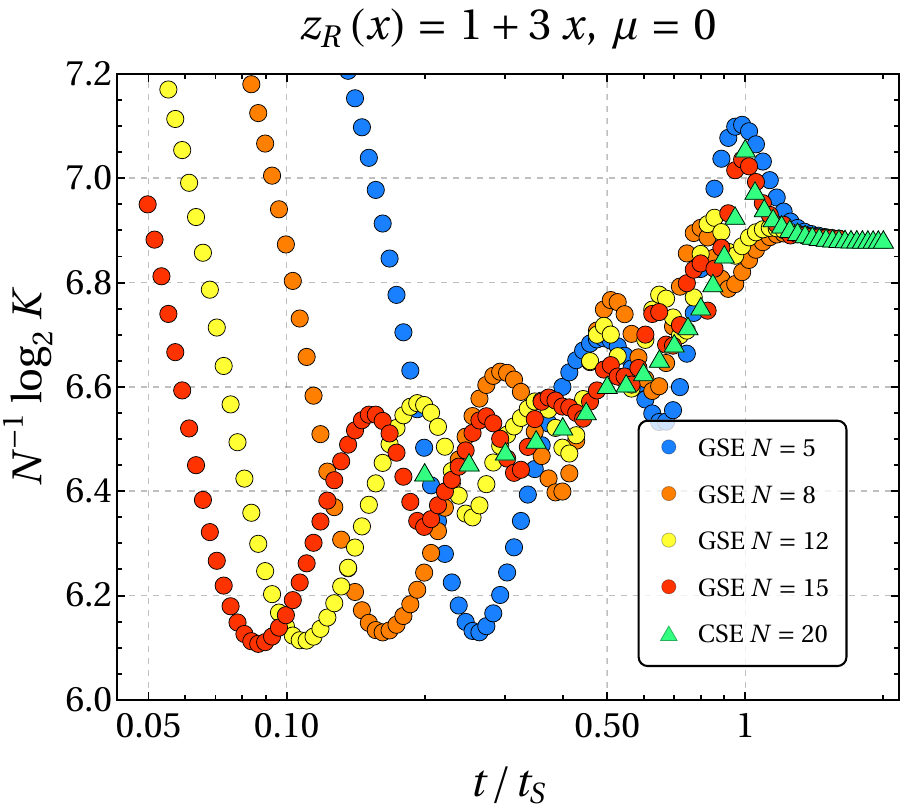}
			\includegraphics[width=0.4\textwidth]{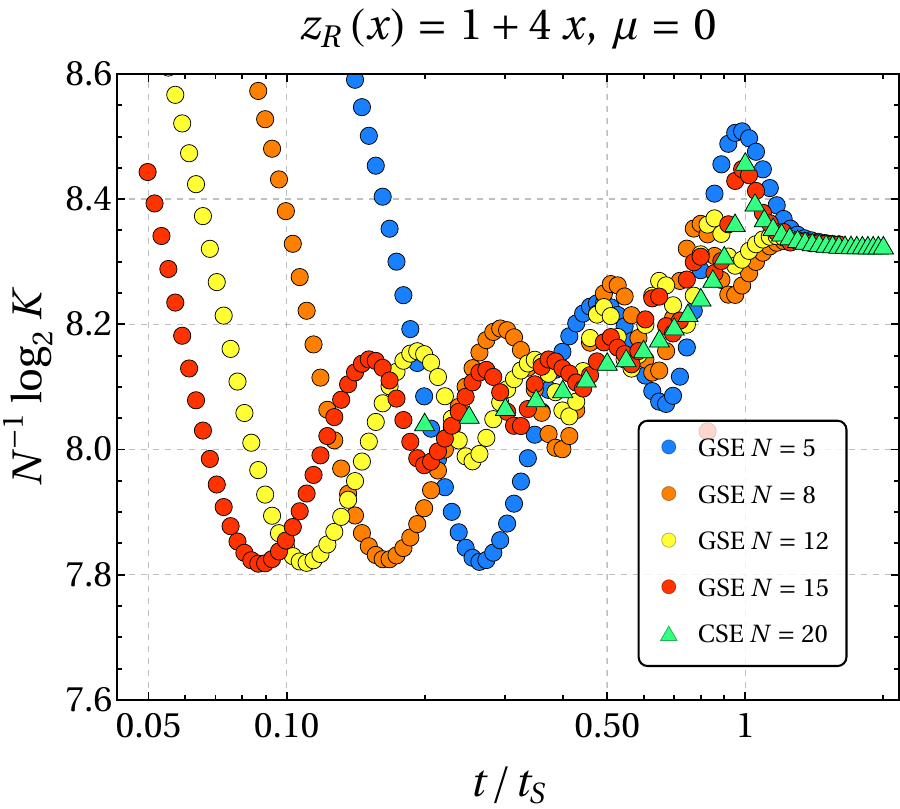}
			\caption{Comparing SFF of CSE and GSE with $z_R(x)=1+mx$. Here data of GUE are generated by Eq.~\eref{eq:GUE-SFF-num} and the data of CUE are obtained by Eq.~\eref{eq:CUE-SFF-recur} with $N=20$. Here scaling time $t_S=1.36 N$ for GUE and $t_S=N$ for CUE. }
			\label{fig:GSEvsCSE}
		\end{center}
	\end{figure}
    \begin{figure}[ht]
    	\begin{center}
    		\includegraphics[width=0.4\textwidth]{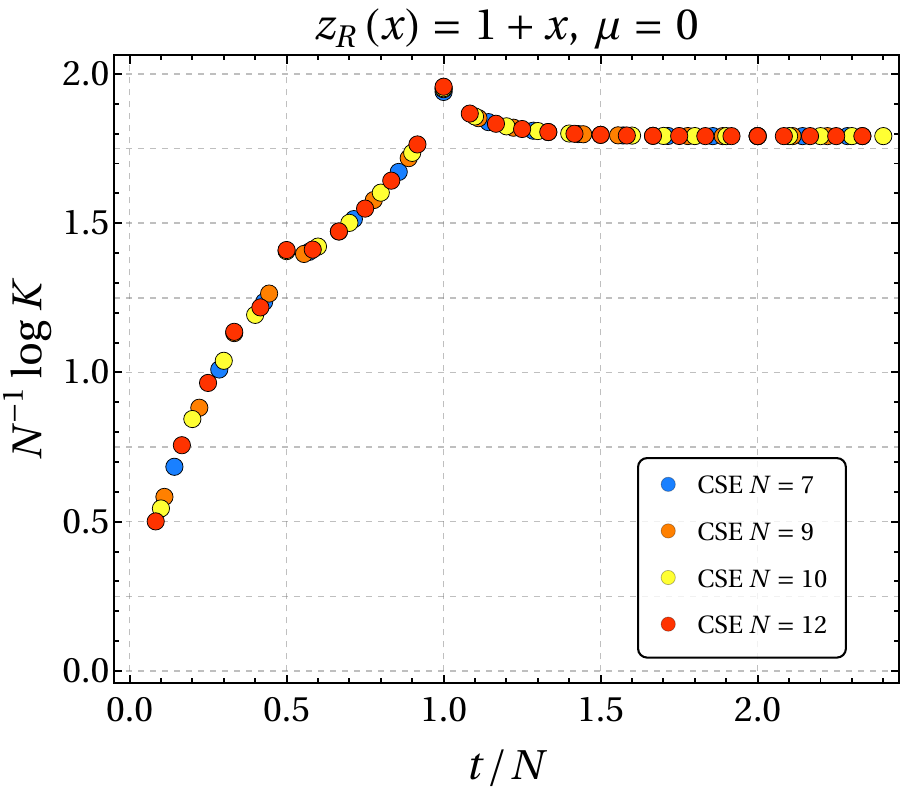}
    		\includegraphics[width=0.4\textwidth]{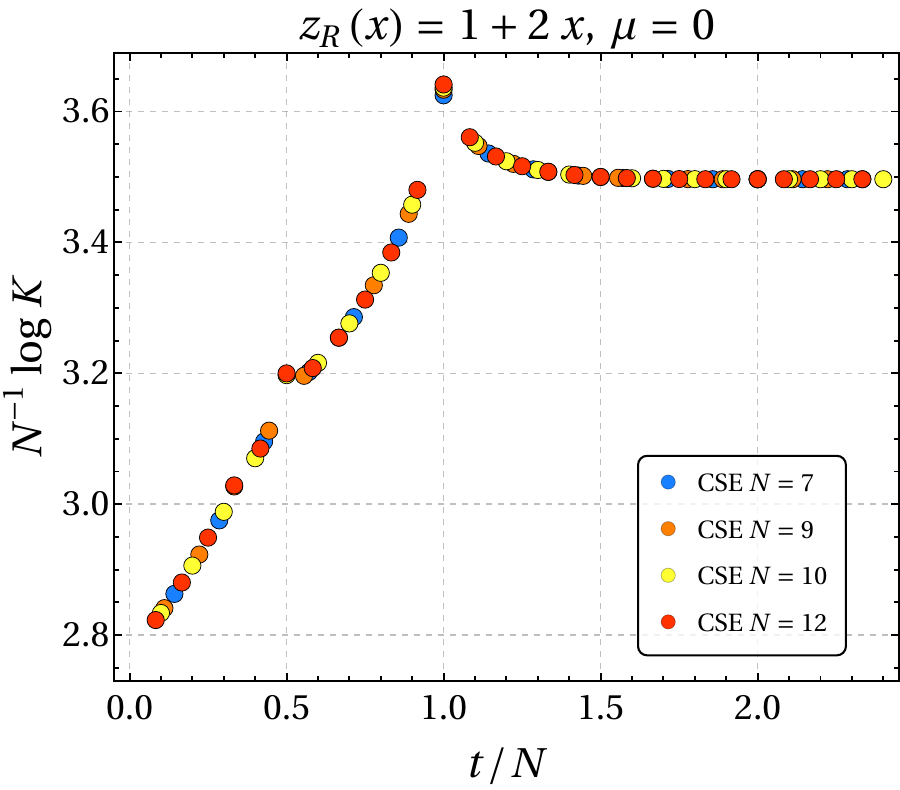}
    		\includegraphics[width=0.4\textwidth]{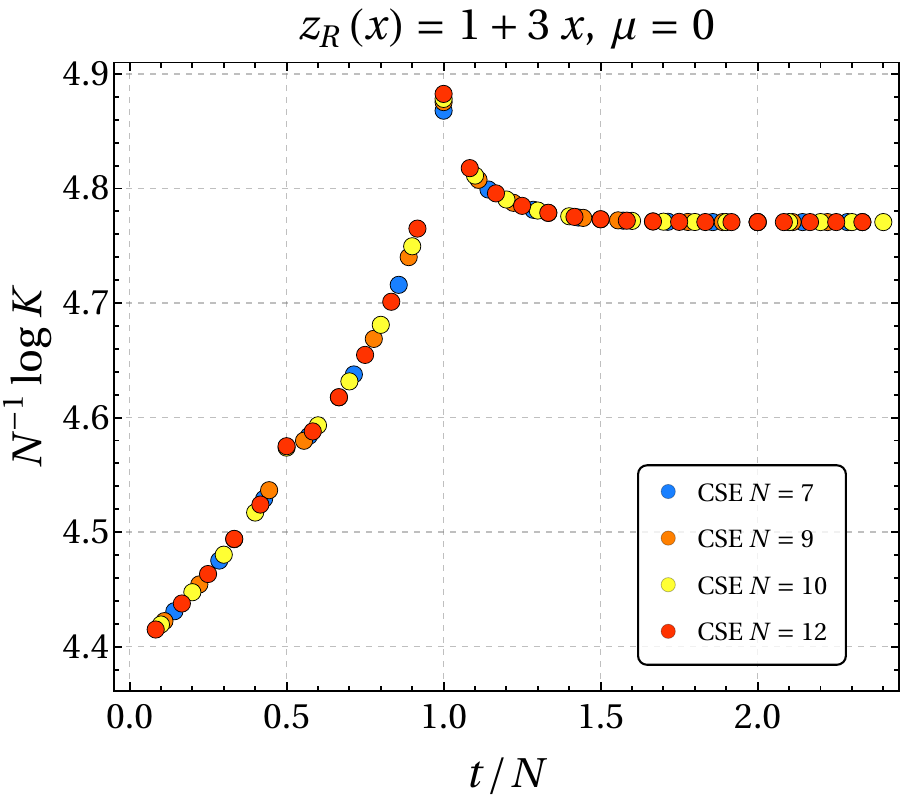}
    		\includegraphics[width=0.4\textwidth]{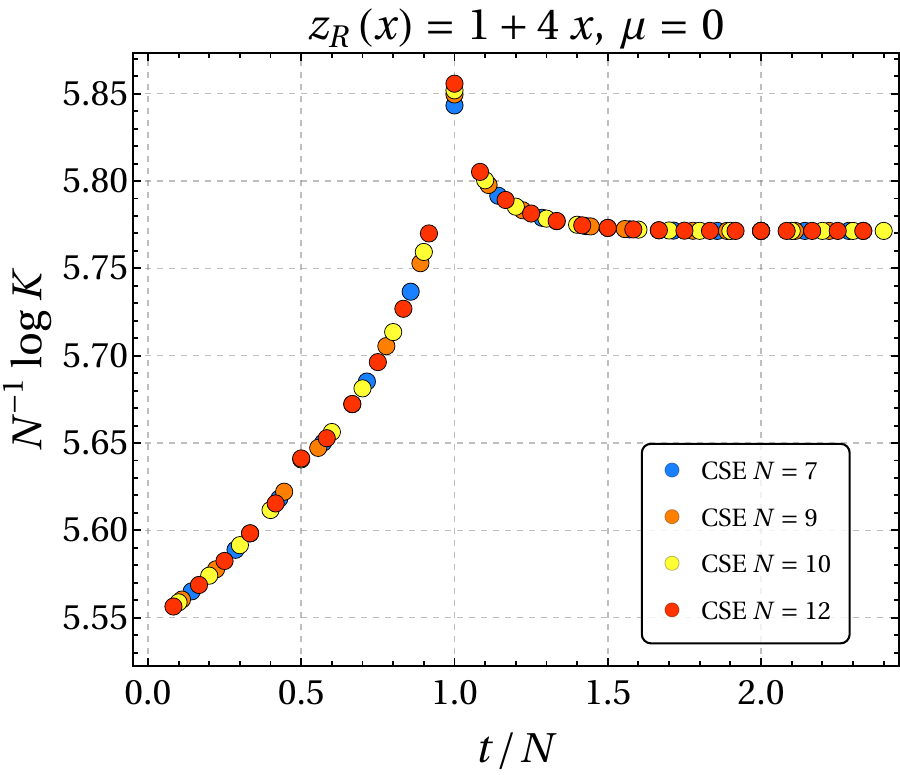}
    		\caption{SFF of CSE coupling with $z_R(x)=1+mx$. Here data of CSE are generated by Eq.~\eref{eq:CSE-Kt-T}.   }
    		\label{fig:CSEvarm}
    	\end{center}
    \end{figure}
    \begin{figure}[ht]
    	\begin{center}
    		\includegraphics[width=0.4\textwidth]{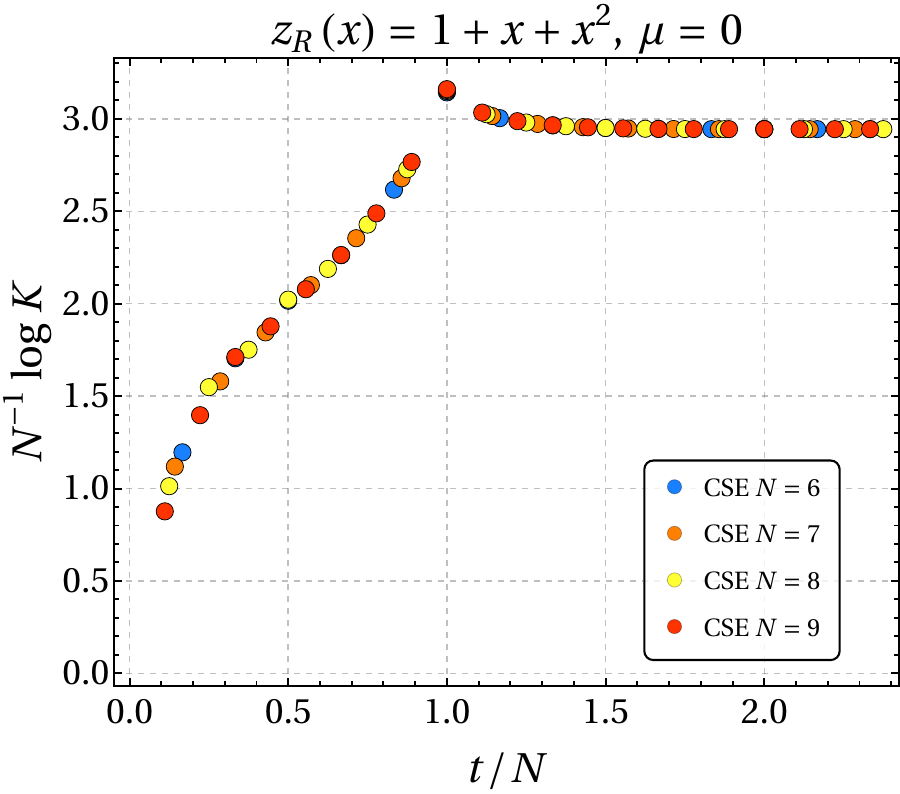}
    		\includegraphics[width=0.4\textwidth]{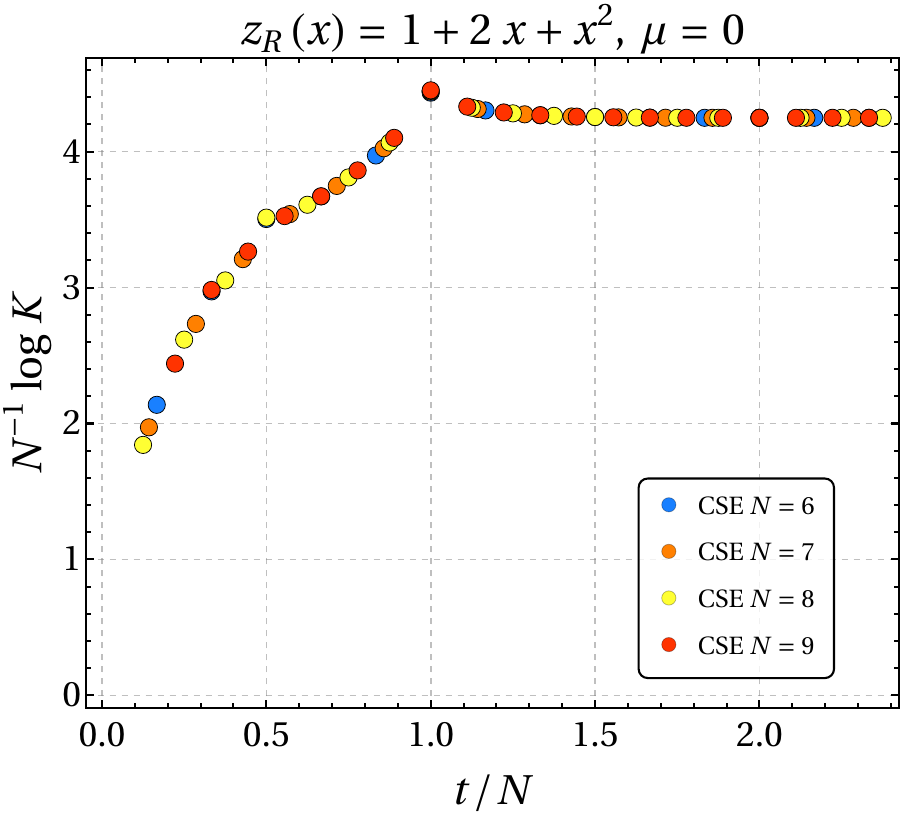}
    		\includegraphics[width=0.4\textwidth]{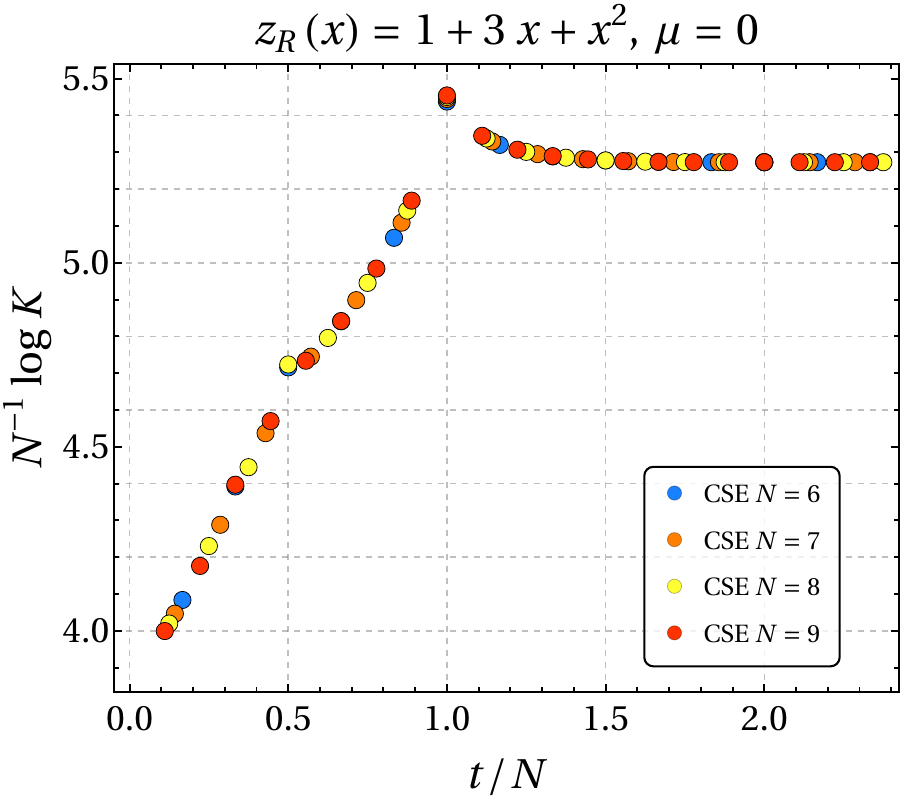}
    		\includegraphics[width=0.4\textwidth]{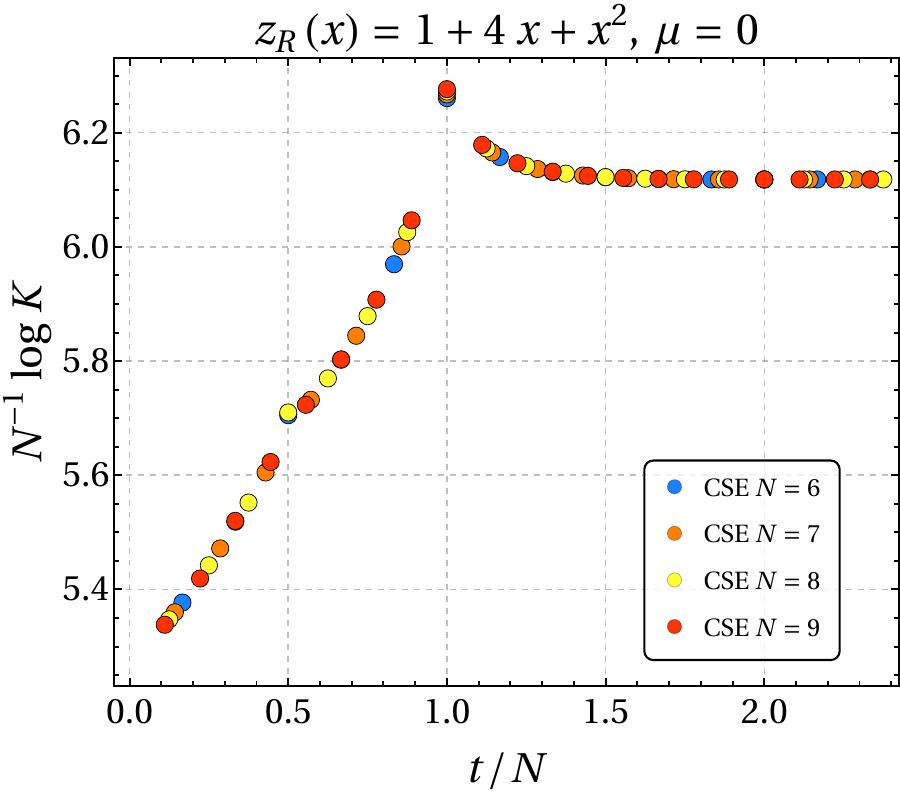}
    		\caption{SFF of CSE coupling with $z_R(x)=1+mx+x^2$. Here data of CSE is generated by Eq.~\eref{eq:CSE-Kt-T}. }
    		\label{fig:CSEexBvarm}
    	\end{center}
    \end{figure}
	As we have seen for the CUE case, such approximation will fail to capture the behavior of SFF when it displays an arc behavior rather than a linear behavior for small $t/N$ region. 
	Unlike GUE case, we do not have an analytical result for GSE $R$-PSYK$_2$, and it is also difficult to see exponential ramp and evaluate its growth rate from numerical data. However, we can use the numerical result of GSE case by assuming the analogy between GSE and CSE. At first, we compare their SFF in Fig.~\ref{fig:GSEvsCSE}, we find they display similar behavior after time scaling, where we choose $t_S=1.36 N$ for GSE. For GUE, exact know the scaling and match the growth rate exactly. But here, we only have a qualitative discussion. 
	
	The exact formula suffers from digital instability for large $N$. The sum of terms in the bracket of Eq.~\eref{eq:CSE-Kt-T} scaled as $\mathcal{O}(e^{-\alpha N}),\alpha>0$, so it is extremely small. Moreover, these terms have different signs, and some of them can be large while they cancel with each other to leave an extremely small number, which leads to a challenge for numerical calculation. For the CUE case, we can only calculate it up to $N=13$. We plot the results for $z_R(x)=1+mx$ and $z_R(x)=1+mx+x^2$ in Fig.~\ref{fig:CSEvarm} and Fig.~\ref{fig:CSEexBvarm} respectively. From the numerical results, we observe that when $m=1$ for Example A and $m=1,2$ for Example B, the slope of the data diverges as $t/N\to 0$. While for other cases, the data near zero exhibit linear growth with slopes that decrease as $m$ increases. These features show complete agreement with the CUE case, although our numerical calculations are limited to moderate system sizes due to computational constraints. For the special case of $m=1$, we refer readers to \cite{Michael_2024} for further details.

	\subsection{COE}
	The COE case will be more complex. To calculate SFF, we again invoke a kernel representation for the $n$-point function $R_n(\theta_n)$. The COE kernel has the matrix representation 
	
	\begin{equation}
		\mathsf{K}_{N,1}(\theta) = \begin{pmatrix} S_N(\theta) & D_N(\theta) \\ J_N(\theta) & S_N(\theta) \end{pmatrix},
	\end{equation}
	where $S_N$ and $D_N$ were defined in Eq.~\eref{eq:SDI-def} and $J$ is given by
	\begin{equation}
		J_N(\theta) = -\frac{1}{2\pi i} \sum_q e^{iq\theta}, 
	\end{equation}
	where $q = \pm(N+1)/2, \pm(N+3)/2, \cdots$, runs over an infinite range. Recall that the Fourier series of $S_N(\theta)$ and $D_N(\theta)$ have a finite number of terms summed over the range $p = -\frac{N-1}{2}, -\frac{N+1}{2}, \cdots, \frac{N-1}{2}$. To avoid writing out these ranges explicitly, it is convenient to develop a unified notation for the kernel matrix elements,
	\begin{equation}
		\mathsf{K}^{\alpha\beta}_{N,1}(\theta) = \frac{i^{\beta-\alpha}}{2\pi} \sum_{p \in \mathbb{Z}+1/2} \eta_{\alpha\beta}(p) p^{\beta-\alpha} e^{ip\theta},
	\end{equation}
	where $p$ is summed over the half-integers, $\alpha, \beta = 0, 1$ index matrix elements of the kernel, and
	
	\begin{equation}
		\eta_{\alpha\beta}(p) = \begin{cases} 
			\Theta \left( |p| - \frac{N}{2} \right) & \alpha = 1 \text{ and } \beta = 0\co \\
			\Theta \left( \frac{N}{2} - |p| \right) & \text{otherwise}\co 
		\end{cases}
	\end{equation}
	so that we have 
	\begin{align}
		\mathcal{T}_{p\alpha,p'\alpha'}=(-1)^{\alpha(1-\alpha')}\eta_{\alpha\alpha'}(p)p^{\alpha-\alpha'}\sum_{\zeta}\mathsf{g}_{\zeta}\delta(p-p'+t\zeta_{j})
	\end{align}
	where we can restrict $|p_{k}|\le\frac{N-1}{2}+tL$, where a similar proof can be found in \cite{Ikeda_2025}. 
	The exact formula is similar to CSE case
	\begin{align}\label{eq:COE-Kt-T}
		\mathcal{K}(t)=D^{2N}\left[1+\sum_{n=1}^N (-1)^{n}\sum_{(\boldsymbol{\lambda},\boldsymbol{m})\vdash n}\prod_{j}\left(\frac{-\text{Tr}\mathcal{T}^{\lambda_{j}}}{2^{2\lambda_{j}+1}\lambda_{j}}\right)^{m_{j}}\right]\ed
	\end{align}
	In this work, we focus primarily on the GUE and CUE cases, and therefore do not present additional numerical results for the COE ensemble.
	\section{Discussion and Outlook}
	\label{sec:discussion}
	In this work, we have systematically investigated the spectral form factor (SFF) of the quadratic $R$-para-particle SYK model coupled with various random matrix ensembles. 
	The circular ensemble approach proved particularly powerful, providing exact results that serve as benchmarks for Gaussian ensemble calculations. While the orthogonal ensemble cases present greater technical challenges, our results establish a unified framework for understanding quantum chaos in quadratic models with generalized statistics. Future directions include extending these methods to interacting systems and exploring connections with other quantum chaotic systems.
	
	As mentioned in the introduction of this paper, the model can be generalized to any joint distribution, allowing us to consider other matrix ensembles. For example, the Wishart (or Laguerre) ensembles
	\begin{equation}
		P_\mathsf{b}(\varepsilon) = c_N^{\mathsf{b}, a} \prod_{i<j} |\varepsilon_i - \varepsilon_j|^\mathsf{b} \prod_{i=1}^N \varepsilon_i^{a-p} e^{-\sum_{i=1}^n \varepsilon_i / 2}, 
	\end{equation}
	or Jacobi ensembles
	\begin{equation}
		P_\mathsf{b}(\varepsilon) = c_N^{\mathsf{b}, a_1, a_2} \prod_{i<j} |\varepsilon_i - \varepsilon_j|^\mathsf{b} \prod_{i=1}^N \varepsilon_i^{a_1 - p} \, (1 - \varepsilon_i)^{a_2 - p}\ed
	\end{equation}
	However, our research aims to understand the essential features of chaotic systems. Currently, there is no exact analytical formula for the SFF of quantum chaotic systems with interactions. This is precisely why we focus on the case of quadratic interactions—the model studied in this work. Our goal is to obtain analytical results while retaining the characteristics of chaotic systems. From this perspective, the Circular Ensemble is sufficiently simple yet non-trivial. For instance, independent and identically distributed $\varepsilon_j$ is simple but trivial, leading to an integrable system rather than chaotic behavior, which is irrelevant to our study of quantum chaos. In contrast, the Circular Ensemble yields an SFF with a ramp and plateau, and for non-integer times, early-time oscillations can also be derived from the literature. Therefore, it is difficult to find a better ensemble for this purpose. We are unaware of any other system that can produce results as robust as those from the Circular Ensemble.
	
	However, the literature's use of the Circular Ensemble provides us with a crucial insight: rather than complicating the problem—such as introducing additional interactions in the SYK model—it may be more fruitful to simplify the model. An example is to consider commuting SYK models, as explored in \cite{Gao:2023gta,Gao:2024lem}. Actually, what we need is a minimal model that preserves the essential physics of interest while remaining sufficiently simple to yield elegant analytical results. 
	
	Thus, we might consider simplifying interacting models systematically. For instance, in the case of the SYK model with $q>2$, a recent simplification strategy involves constructing it using mutually commuting operators. Alternatively, we could circumvent the challenging diagonalization step entirely by postulating the energy spectrum's functional form from the outset. By choosing an appropriate disorder distribution, we may then reproduce a generic linearly increasing ramp in the chaotic SFF. Here, the spectrum could be a linear function of particle number operators and random variables; without disorder, this would remain a free model. However, imposing a nontrivial joint distribution on these random variables could induce chaotic behavior. 
	
    To mimic interactions more realistically, we might further generalize the spectrum to quadratic (or higher-order) functions of random variables and particle number operators. For example,
    \begin{equation}
    	H = \frac{1}{N}\sum_{i,j}\varepsilon_{i}\varepsilon_{j}f(n_{i}n_{j})\co
    \end{equation}
    where $f(n_i,n_j)$ can be an arbitrary function of occupation numbers $n_i,n_j$. For simplicity, these variables $\varepsilon_j$ could be chosen as independent and identically distributed. These directions will form the basis of our future investigations.
	
	\section*{Acknowledgments}
	I am grateful to Yingyu Yang, Yanyuan Li and Xingpao Suo for their insightful discussions. T.L. acknowledges support from the National Natural Science Foundation of China (Grant No. 12175237).

	\bibliographystyle{JHEP}
	\bibliography{ref}
\end{document}